\documentclass{eas}
\usepackage{graphicx}
\usepackage{astron}
\begin{document}

\title{Observational Evidence for Tidal Interaction \\ in Close Binary Systems} 
\runningtitle{Observation of Tidal Interaction in Close Binary Systems}
\author{Tsevi Mazeh}\address{School of Physics of Astronomy, Sackler
  Faculty of Exact Sciences, Tel Aviv University}

\begin{abstract}

This paper reviews the rich corpus of observational evidence for tidal
effects, mostly based on photometric and radial-velocity measurements.
This is done in a period when the study of binaries is being
revolutionized by large-scaled photometric surveys that are detecting
many thousands of new binaries and tens of extrasolar planets.

We begin by examining the short-term effects, such as ellipsoidal
variability and apsidal motion.  We next turn to the long-term
effects, of which circularization was studied the most: a transition
period between circular and eccentric orbits has been derived for
eight coeval samples of binaries.  The study of synchronization and
spin-orbit alignment is less advanced. As binaries are supposed to
reach synchronization before circularization, one can expect finding
eccentric binaries in pseudo-synchronization state, the evidence for
which is reviewed. We also discuss synchronization in PMS and young
stars, and compare the emerging timescale with the circularization
timescale.

We next examine the tidal interaction in close binaries that are
orbited by a third distant companion, and review the effect of
pumping the binary eccentricity by the third star. We elaborate on the
impact of the pumped eccentricity on the tidal evolution of close
binaries residing in triple systems, which may shrink the binary separation.

Finally we consider the extrasolar planets and the observational
evidence for tidal interaction with their parent stars. This includes
a mechanism that can induce radial drift of short-period planets,
either inward or outward, depending on the planetary radial position
relative to the corotation radius.  Another effect is the
circularization of planetary orbits, the evidence for which can be
found in eccentricity-versus-period plot of the planets already known.

Whenever possible, the paper attempts to address the possible
confrontation between theory and observations, and to point out
noteworthy cases and observations that can be performed in the future
and may shed some light on the key questions that remain open.

\end{abstract}
\maketitle



%
\section{Introduction}
%

Stars in close binary systems are subject to mutual tidal forces that
distort their stellar shape, breaking their spherical and axial
symmetry, which leads to different observational effects.  This paper
reviews the rich corpus of observational evidence for the tidal
effects, which has been accumulated mostly by photometric and
radial-velocity studies of short-period binaries.

One can divide the observational effects of the stellar distortion
caused by tidal interaction into three classes:

\begin{itemize}

\item 
Direct observations of the distorted shape of the two components of
the binary system. One example is the ellipsoidal periodic modulation
of the photometric intensity of the binary with the orbital period,
caused by the rotation of the stars with the ellipsoidal shape.

\item
Deviation of the orbital motion of the binary from pure Keplerian
orbit. This is caused by departures of the gravitational attraction
from the inverse-square law, due to the distortion of the two stars
from spherical symmetry. One example is the famous apsidal motion
observed in many eccentric eclipsing binaries.

\item
Long timescale evolution of the binary orbital elements and the
stellar rotation, induced by the tidal interaction.  The long-term
interaction will take place as long as the tides raised by the mutual
interaction vary in size, shape and location on the surface of the two
components. The tides do not vary only when

\begin{itemize}

\item 
the binary motion is circularized,

\item
the stellar rotation is synchronized with the binary orbital
motion, and

\item
the stellar rotation axis is aligned with the normal to the orbital plane
of motion.

\end{itemize}
Observational evidence for this long-term evolution can be found, for
example, through the circular orbits of short-period binaries.

\end{itemize}

The timescales of the three classes are substantially different. The
ellipsoidal effect varies with the orbital period, which is of the
order of a few days. The timescale of the deviations of the binary
motion from the Keplerian orbit is in the range between tens and
thousands of years, whereas the last class timescale is of the order
of millions or billions of years. Therefore, we can witness close
binaries for which the effects of the first and the second class are
in action, while observation of the third class of effects can be made
only retroactively, in recognition of the results of a long timescale
processes.

Observing one or more of these effects allows us to learn about the
response of each star to the tidal forces exerted by its
companion. This response depends on the stellar internal structure, and therefore observations of these effects can provide us
with a unique opportunity to study that structure in a
manner which is not possible with single stars.

In the last few decades numerous observations of the tidal interaction in
close binaries have been accumulated.  In many cases, observations of
binaries performed for other reasons could not have been accounted for
without the tidal effects. One example is the lightcurves of eclipsing
binaries, which are usually observed in order to derive the
geometrical elements of the systems. However, the lightcurves of these
close systems can not be fully understood without taking into account
the distortion of the two stars. In other cases, special observational
effort has been made in order to detect some evidence for tidal
effects. One example is the search for the circularization cutoff
period in samples of coeval binaries. Both types of cases give us an
opportunity to confront the tidal theory, based on stellar structure
and evolution, with the observations. This paper will review both
types of observations.

Sections 2 \& 3 review the observational evidence for the
ellipsoidal modulation and the apsidal motion of eccentric binaries.
Section 4, 5, and 6 discuss the evidence for long-term
circularization, synchronization and alignment of short-period
binaries. Section 7 covers the tidal interaction in triple systems,
when the third distant companion can have an impact on the tidal
evolution of the close binary. 

The study of tidal interaction between the recently discovered
extrasolar planets and their parent stars has gained rapid momentum
since 1995, following the discovery of the first confirmed extrasolar
planet. Section 8 reviews some aspects of this fascinating new field.

This paper has been written at a juncture when the study of close
binaries has been undergoing a transitional phase, the result of
large-scaled photometric surveys which are starting to produce large
samples of eclipsing binaries. In acknowledgment of the new era,
Section 9 reviews some of the recent results from two of these
surveys. Finally, Section 10 summarizes the previous sections and
discusses some aspects of future studies of tidal interaction in close
binaries.

This paper is too short and my knowledge too incomplete to cover all
the studies in the different aspects of the subject. I made a great
effort, though, to cover in each of the sections at least the open
questions in that subfield. I apologize to those whose work was not
included in this review.  I have made an effort to keep the notation
of each section consistent with the other sections of this
review. This forced me to adopt some notations that are not so common
in the different fields covered by this paper. Each section was
written as a separate piece of study, so many definitions were
repeated throughout this paper.

Finally, I hope that this review will be of some use to young
researchers in the field. It would fulfill my highest expectations if
some research is initiated as a result of reading this review and
considering some of questions that it raises.

%
\section{Ellipsoidal effect}
%

Ellipsoidal variables are close binary systems whose lightcurves show
periodic variations due to the distortion of the components caused by
their mutual attraction. Usually, this term is used (e.g., Morris \&
Naftilan \nocite{mornaf93} 1993) to denote binaries whose orbital
inclinations are not large enough to render them eclipsing systems,
although the effect is seen in eclipsing binaries as well.

When calculating the ellipsoidal effect (e.g., Kopal 1959, Morris
\nocite{morris1985}1985, Morris \& Naftilan 1993) one usually assumes
that the two stars are in equilibrium in the binary rotating reference
frame, which means that the stars are synchronized and aligned and
that the orbit is circularized.  Models of stellar evolution and
circularization and synchronization (e.g., Zahn 1977) have shown that
for binaries with periods less than a few days these conditions are
reached on a much shorter timescale than the lifetimes of the
stars. Extensive observational evidence supports this theoretical
result (see Sections 4--6).  For longer periods, the ellipsoidal
effect is small anyway.

To derive the ellipsoidal effect, Morris \& Naftilan (1993) expanded
the periodic variation into discrete Fourier series with terms that
depend on the ratio $R_*/a$, where $R_*$ is the stellar radius and $a$
is the semi-major axis of the binary orbit. Assuming $R_*/a$ is small,
the leading term to present the variation of the {\it primary}, with a
radius $R_1$, of the
order of $(R_1/a)^3$, has a semi-amplitude of:
\begin{equation} 
\left( \frac{\Delta F_1}{F_1}\right)_{\mathrm{ellips}} =
0.15\frac{(15+u_1)(1+g_1)}{3-u_1}\frac{M_2}{M_1}
\left(\frac{R_1}{a}\right)^3
\ \sin^2 i \ .
\end{equation} 
In the above expression, $g_1$ is the gravity darkening coefficient
of the primary and $u_1$ is its limb-darkening coefficient, and $i$ is
the orbital inclination. 

For a period of a few days and in main-sequence stars, the amplitude
of the ellipsoidal effect is a few percent, and therefore the effect
can be easily measured. For a spectroscopic binary, known by its
radial-velocity modulation, measuring the amplitude of the ellipsoidal
effect can substantially add to our knowledge of the system, as this
amplitude can add some constraints on the stellar radius, mass and the
binary inclination. If we can estimate the stellar radius from its
spectral type and the semi-major axis of the binary from its period,
we can estimate the inclination of the binary, which is a crucial
parameter for accurately obtaining the stellar masses of double-lined
spectroscopic binaries.

Observational astronomy is always complex, and the ellipsoidal effect
is no exception. First, when calculating the expected amplitude of the
effect, one must take into account the variation of the secondary star
as well. In fact, as the intensities of the two stars are both
modulated with half the orbital period, the observed effect is the
weighted average of the two stars. Therefore, deriving the inclination
from the ellipsoidal amplitude is not a trivial matter and might
necessitate observations in a few wavelengths. Second, other stellar
modulations might also be present, like in the case of the $\delta$
Scuti star XX Pyx (Aerts \etal\ 2002)\nocite{aerts2002}.  Fortunately,
the ellipsoidal effect has a clear periodicity of half the orbital
period, which is different, for example, from the stellar spot
modulation that varies, for a synchronized system, with the orbital
period.

\subsection{Interesting systems: Black holes in X-ray binaries}

One field in which the ellipsoidal variation was found to be critical is
the study of X-ray binaries, which are close binaries with a compact
object --- a white dwarf, a neutron star or a black hole, together
with a main-sequence or a giant optical companion. The X-ray
luminosity, generated around the compact object, is fuelled by mass
transfer from the companion onto the compact object. In many of these
systems, the companion fills its Roche lobe and therefore its shape is
substantially distorted and the corresponding ellipsoidal effect is
relatively large.

In almost all the X-ray binaries, the ellipsoidal effect is crucial to
putting strong constraints on the orbital inclination and the mass of
the compact object.  The mass of the compact object is the only way to
identify stellar black holes (=BH), and therefore the ellipsoidal
variability is an essential step in studying these general-relativity
objects (e.g., Orosz 2003\nocite{orosz2003}). Although one has to take
into account the unknown contribution of the accretion disc to the
luminosity of the system, careful analysis can give some constraints
on the mass of the compact object (e.g., Charles \& Coe
2006\nocite{charles2006}). Two such examples are V616 Mon (e.g.,
Harrison \etal\ \nocite{harrison2007}2007) and XTE J118+480 (Gelino
\etal\ \nocite{gelino2006}2006).

An interesting case is M33 X-7, the first stellar BH found in the
neighbouring galaxy M33 (Pietsch \etal\ \nocite{pietsch06}2006).  With
a period of 3.45 days, the ellipsoidal amplitude was measured to be of
about 4\% (Shporer \etal\ \nocite{shporer07}2007), and contributed to
the derivation of a mass of $15.65 \pm 1.45$ $M_{\odot}$ (Orosz \etal\
\nocite{orosz2007}2007).

%
\section{Apsidal motion}
%

Apsidal motion, the precession of the orbital ellipse in the binary
plane of motion, can be measured by the detection of a small change of
$\omega$ --- the angle denoting the direction of the line of apsides
in the orbital plane. In a Keplerian motion the ellipse and its
orientation are constants of motion, and only deviation from
Newtonian inverse-square law attraction can cause the ellipse to
precess. The deviation results from the asymmetrical stellar shape,
caused by the tidal forces of its companion. The precession period is
longer than the orbital period by many orders, and therefore one can
first derive the osculating orbital parameters, ignoring the apsidal
precession, and only afterward search for variations of $\omega$ over a
longer timescale.

The classical theoretical derivation of the precession rate ignores
the general relativity effect and assumes spin-orbit alignment of the
two stars (see discussion of the two effects in Subsections 3.1 and 3.2). 
To calculate theoretically the expected rate of apsidal motion one has
first to derive the stellar shape induced by the tides, and then
to calculate the resulting advance of the line of apsides. For an
historical review of the theoretical calculation which started back in
1928 by Russell \nocite{russell1928} (1928), see Claret \& Gim{\'e}nez
\nocite{claret1993} (1993). The apsidal precession period due {\it
only} to the primary star, $U_{tidal,1}$, taking into account only the
linear part of the stellar quadruple moment and up to the second
order in the eccentricity $e$, is given by the formula:
\begin{equation}
\frac{P}{U_{tidal,1}} \simeq k_{2,1}\left(\frac{R_1}{a}\right)^5 
\left[15f_2(e)q+\left(\frac{\Omega_{1,rot}}{n_{orbit}}\right)^2 
(1+q) \right]
\end{equation}
where $P$ is the orbital period, $q$ is the mass ratio --- $M_2/M_1$,
where $M_1$ is the primary mass and $M_2$ is the secondary mass,
$\Omega_{1,rot}$ is the {\it rotational} frequency of the primary,
$n_{orbit}=2\pi/P$ is the averaged orbital frequency, $R_1$ is the radius of
the primary, $a$ is the orbital semi-major axis, and $f_2(e) \simeq
(1+\frac{3}{2}e^2)$ is a function of $e$ of order unity. The parameter
$k_{2,1}$ is the second-order coefficient of the internal structure of
the primary, and reflects the stellar radial concentration (e.g.,
Cowling 1938\nocite{cowling1938}; Sterne \nocite{sterne1939a}
1939a; b; c\nocite{sterne1939b}\nocite{sterne1939c}; Kopal
\nocite{kopal1959}1959). For infinite degree of concentration of the
primary (mass point) $k_{2,1}$ becomes zero, while for a homogeneous
configuration $k_{2,1}=0.75$ (e.g., Claret \& Gim{\'e}nez
\nocite{claret1993} 1993). Typical models yield $k_{2,1}$ of the order
of 0.01--0.001, depending on the stellar mass and age. The precession
is in the direction of the orbital motion, and therefore
$\dot{\omega}$ is always positive. Such a precession is sometimes
termed direct precession.
 
In most systems one must also take into account the secondary
contribution to the precession, and therefore one uses $\bar{k}_2$ ---
some weighted average of $k_{2,1}$ and $k_{2,2}$, the latter being the
structure parameter of the secondary (e.g., Claret
\nocite{claret1999}1999).
 
Apsidal motion can be measured with relative accuracy in eccentric {\it
eclipsing} binaries, as $e$ and $\omega$ determine the time interval
between the primary and secondary minima (e.g., Guinan \& Maloney
\nocite{guinan1985} 1985). When the time interval is substantially
different from half the orbital period, small changes of $\omega$ can
be detected. Observational determination of timing of minima, performed
over a baseline of the order of ten years, can reveal a precession of a
degree or two, indicating a precession period of thousands of years.

For double-lined {\it spectroscopic eclipsing} binaries, where the
masses and radii of the two stars can be derived from the lightcurve
and the radial-velocity curves, observation of the
apsidal precession period can directly tell us the value of
$\bar{k}_2$. Therefore, the apsidal motion is a way to confront the
theory of stellar structure with observations (e.g., Gim{\'e}nez et
al. 1987\nocite{gimenez1987}; Gim{\'e}nez\nocite{gimenez1990} 1990). There
are very few contemporary observations that can directly challenge
the stellar structure theory without relying on stellar
evolution. Solar neutrino experiment (e.g., Bahcall \& Ulrich
\nocite{bahcall1988}1988) and stellar seismology (e.g., Dziembowski \&
Pamyatnykh \nocite{dziembowski1991}1991) are two of the very few other
examples. This is why the apsidal motion, although limited to
eclipsing binaries with short periods, is of fundamental importance to
stellar astrophysics. For a compilation of the binaries with measured
apsidal motion up to their time see Petrova \& Orlov
(\nocite{petrova1999}1999). Very recently Bulut \& Demircan
(\nocite{bulut2007}2007) published a new catalog of 124 eclipsing
binaries, including the pertinent observed apsidal motions.

Claret \& Gim{\'e}nez \nocite{claret1993} (1993; see also Claret
\nocite{claret1999} 1999; \nocite{claret2007} 2007) carefully
considered the observational result of 14 double-lined eclipsing
binaries with known absolute dimensions and accurately observed
apsidal motion rates. They also included in their study an additional
ten systems with lower quality data. Although there have been
arguments for some inconsistencies between the old stellar models and
the observed apsidal motion, Claret \& Gim{\'e}nez found that the
observations are consistent with their modern stellar models (see also
Claret \& Willems \nocite{claret2002} 2002 and Petrova \&
Orlov \nocite{petrova2002}2002; 2003). As in the case of the solar
neutrino (Bahcall \etal\ 2001\nocite{bahcall2001}), the consistency
between the theory and observation of the apsidal motion for most of
the binaries should be considered one of the great achievements of
stellar astrophysics.

Apsidal motion can still refine our understanding of stellar
interiors. This is especially true of stars with convective cores,
where the present models include a somewhat arbitrarily chosen value of
the overshooting parameter (e.g., Claret \& Willems 2002). The value
of this parameter can be deduced from the apsidal motion precession
period, as was done in the study of V380 Cyg (Guinan \etal\
\nocite{guinan2000}2000), which consists of two B stars.

Given the agreement between theory of stellar structure and 
observations, we can now reverse the reasoning and use the observed
tidal apsidal precession to derive the masses of the systems for
spectroscopic binaries that do not show any eclipse (Jeffery
\nocite{jeffery1984}1984). For those systems, detection of the tidal
precession can provide missing information about the orbital
inclination, information that can be found in eclipsing binaries from
the depth, shape and duration of the eclipse. Although the measurement
of the precession in spectroscopic binaries is more difficult than in
eclipsing binaries, Benevenuto \etal\ (\nocite{benenuto2002}2002) used
this method to derive the masses of the two O-stars in the HD 93205
binary system.

\subsection{Relativistic Precession}

Deviation from inverse-square law gravitational attraction in binary
systems appears also because of a general relativistic (=GR) effect that
in some cases can not be neglected. The resulting direct apsidal
precession amounts to

\begin{equation}
U_{GR} = 1800 (1-e^2) \left(\frac{P}{\rm{day}}  \right)^{5/3}
 \left(\frac{M_1+M_2}{M_{\odot}}  \right)^{-2/3} \,{\rm yrs} \ .
\end{equation}
(e.g., Weinberg 1972\nocite{weinberg1972}). 
The combined precession period, $U_{tot}$, is
\begin{equation}
\frac{1}{U_{tot}}=\frac{1}{U_{GR}}+\frac{1}{U_{tidal}} \ .
\end{equation}
Therefore, in order to derive the tidal precession from any observed
precession we have, in principle, to subtract the GR
effect. Fortunately, the GR precession can be predicted accurately, as
it depends only on orbital parameters and does not depend on the
stellar radius and internal structure.

In order to compare the two precessions, we note that although for
constant $R_1/a$ the apsidal motion goes linearly with $P$, for
constant primary stellar radius $R_1$ we get that

\begin{equation}
 {1 \over U_{tidal}} \propto P^{-13/3} \ \ \ {\rm while} \ \ \ {1 \over U_{GR}}\propto P^{-5/3} \ . 
\end{equation}
Therefore, for binaries with long enough periods, the tidal
precession becomes dominant and the GR precession can be neglected
(see Figure~\ref{apsidal_motion_LMC} and the relevant discussion). 

\subsection{The effect of spin-orbit misalignment on the precession rate}
\label{DI_Her}

Another effect that can change the apsidal precession rate is
spin-orbit misalignment of one or two of the stars in the binary
system. One possible example for this effect is DI Her, which is one
of the eclipsing binaries that has presented a persistent challenge to
the theory of apsidal precession (Martinov \& Khaliullin
\nocite{martynov1980} 1980) for more than 25 years. The system
consists of two bright B stars orbiting each other with a period of
10.55 days and eccentricity of $e=0.49$. For an historic account of
the study of the system see Guinan \& Maloney
(\nocite{guinan1985}1985, hereafter GM85).

Modern value for the apsidal motion of DI Her was observed (GM85) to
be $55,400$ yrs.  However, the stellar models, which rely on the
observed stellar radii and rotations, yielded a much shorter tidal
precession period --- of the order of 20,000 yrs. The inconsistency
with the theory got worse, as the GR precession period, which does not
depend on any stellar model, was estimated to be even shorter ---
$15,400$ yrs. Together, the two effects amounted to a precession period
as short as 8,400 yrs, almost one order of magnitude shorter than the
observed apsidal period. Such a confrontation between theory and
observations can be very fruitful. It can indicate either that the
theory is wrong, or that we are ignoring effects that are important
for the system.

Shakura (\nocite{shakura1985}1985; see also Company \etal\
\nocite{company1988}1988) suggested one possible way to 'save the
phenomenon' of DI Her, by assuming the rotation axis of at least one
of the two stars is not aligned with the orbital angular momentum of
the system.  In such a case, the companion's gravitational force
exerted on the equatorial stellar rotational bulge induces a {\it
counter} precession of the orbit, the rate of which depends on the
angle of misalignment and the size of the equatorial bulge. The net
result of the misalignment could be a substantial reduction of the
apsidal precession rate. Reisenberger \& Guinan
(\nocite{reisenberger1989}1989), who studied Shakura's idea in depth,
concluded that in order to account for the observed slow precession of
DI Her one has to assume a misalignment of about $50^{\circ}$. 

DI Her is not the only binary for which the apsidal motion might
indicate some spin-orbit misalignment. Petrova \& Orlov
(\nocite{petrova2003}2003) discussed the possible misalignment effect in
broader context, and considered a few more interesting systems.

The alignment of the rotational axes of the stars with the angular
momentum of their binary orbit is discussed in detail in
Section~\ref{section_alignment}. The conjecture that the stellar spin
axes are not aligned with the binary motion involves two interesting
theoretical assumptions. First, if this idea is true for DI Her and
the other systems, it means that these binaries were formed with
misaligned stellar rotation. Second, the present misalignment means
that the long-term tidal dissipation processes did not have enough
time to align the stellar spins (see Section 6 below). In fact,
Reisenberger \& Guinan (1989) estimated that the lifetime of the two
components of DI Her is substantially shorter than the alignment
timescale, and therefore one would expect DI Her to retain its
original spin-orbit misalignment.

Spin-orbit misalignment can cause two additional interesting
phenomena, in addition to the counter apsidal precession. The binary
interaction causes the stellar rotation axis {\it and} the orbital
orbit to precess around the total angular momentum of the system with
the same period. Such a forced precession will change the stellar spin
relative to our line of sight, resulting in a change of the stellar
rotational broadening of the two stars.  It might also change slightly
the inclination angle of the binary, which modifies the eclipse
parameters. Reisenberger \& Guinan (\nocite{reisenberger1989}1989)
followed Shakura's suggestion and tried to find observational hints
for these two modulations, but the evidence they have collected was
inconclusive.  Since almost 20 years have passed since the study
of Reisenberger \& Guinan, an observational revisit of the system
might yield more reliable evidence that can either confirm or refute the
orbital and stellar precession, as the time baseline of the
inclination measurements will be then quite longer.

We note that the advance of the observational techniques since the
mid-1980's renders the Shakura conjecture for DI Her easily
observable. During the eclipse one can observe the Rossiter-McLaughlin
effect (Rossiter \nocite{rossiter1924} 1924; McLaughlin
\nocite{mclaughlin1924} 1924 --- see Section~\ref{rm_effect} for an
extensive discussion), by which the stellar absorption lines display
systematic deviation from their normal rotational broadening profile,
because some parts of the stellar surface are being eclipsed. As
Section~\ref{V1143_RM} points out, such observations were recently
obtained for V1143 Cyg (Albrecht \etal\ 2007). It would be of great
interest to confirm Shakura's idea, particularly because of its
implications for binary formation, as well as because competing ideas
to solve the mystery of DI Her have been discussed in the literature
(Claret \nocite{claret1998}1998).

%
\section{Circularization}
%

We move now
to consider the observational evidence for long-term dissipative
processes taking place in short-period binaries. This section will
consider the evidence for circularization, while the next two sections
will review the evidence for synchronization and alignment.

It is not easy to find observational evidence for the circularization
processes, as we are not able to follow any binary that goes through
the long-term circularization. Instead, we can only observe circular
binaries that we {\it guess} were formed with eccentric
orbits. However, like every retrospect line of evidence, such
observations are open to other interpretations, and the conclusions are
not unquestionable.

On the theoretical side, the calculations of the circularization
processes and their effects are more complicated than those of the
ellipsoidal modulation and the precession period. This is so because
in the latter two processes one assumes an immediate stellar response
to the tidal forces, while the derivation of circularization timescale
involves a response {\it lag} between the tidal forces and the stellar
shape (e.g., Zahn 1966). As we shall see, this entails serious
disagreement between the theoreticians, which makes the observational
evidence even more important.

To first approximation, the tidal interaction between the two
components of a binary reduces the orbital eccentricity $e$ such that
\begin{equation}
{de\over{dt}}=-Ce \, ,
\end{equation}
where the factor $C$ depends on the orbital separation, the internal structure of the two stars and
their rotation. As $C$ varies very slowly, usually on the timescale of
the stellar lifetime, we can assume in many cases that $C$ is
constant. Therefore, the eccentricity decays exponentially, and we can
define the circularization timescale $\tau_{circ}$ with the equation:

\begin{equation}
-{d\ln e\over{dt}}={1\over{\tau_{circ}}}
\label{eq_circ}
\end{equation} 
(Zahn 1975).  This equation implies that within a few $\tau_{circ}$
the binary motion assumes a nearly circular orbit, independently of its
primordial eccentricity. We therefore expect any binary older than
three or four times its $\tau_{circ}$ to become circular. 

As shown by the seminal work of Zahn (1975), the circularization
timescale is very sensitive to the orbital semi-major axis $a$:

\begin{equation}
\tau_{circ} \propto \left\{
\begin{array}{c c}
\left(a/R_1\right)^{8}  
& \ \ \mbox{for stars with } convective \mbox{ envelopes} \\
\ \ \ \left(a/R_1\right)^{13/2}  
& \mbox{for stars with } radiative \mbox{ envelopes} \\
\end{array} \right. 
\label{circularization_eq}
\end{equation} 
where $R_1$ is the radius of the primary and $a$ is the binary
semi-major axis, assuming the tidal dissipation in the secondary can be
neglected. Determination of the eccentricities of short-period
binaries, for which $\tau_{circ}$ is necessarily short, can therefore
support the circularization idea observationally, {\it if} we find
that all or at least most of the short-period binaries have circular
orbits. Furthermore, considering a sample of {\it coeval} binaries,
the eccentricity as a function of the binary period can tell us how
the relevant $\tau_{circ}$ depends on the binary separation and the
stellar structure.

\subsection{The SB9 sample of spectroscopic binaries}

One indirect evidence for long-term circularization processes can
be obtained by considering the eccentricities of all known
spectroscopic binaries as a function of their orbital period.  To do
that we use the 2751 systems compiled by the official IAU catalog of
{\it spectroscopic} binaries --- SB9 (Pourbaix \etal\
\nocite{pourbaix2005}2005), as of August 2007. The sample of the
catalog is extremely inhomogeneous and includes stars of different
spectral types and ages. In addition, the binaries compiled
were discovered by different surveys with different observational
instruments,  rendering their discovery thresholds
different as well. Furthermore, the compilation of the catalog is in progress,
and therefore the catalog is not yet complete. Nevertheless, the catalog
presents the largest sample of known spectroscopic binaries, and
consequently we use it as a valuable source to study some features of
the population of binaries in general, and short-period ones in particular.

\begin{figure}
\includegraphics[width=13cm]{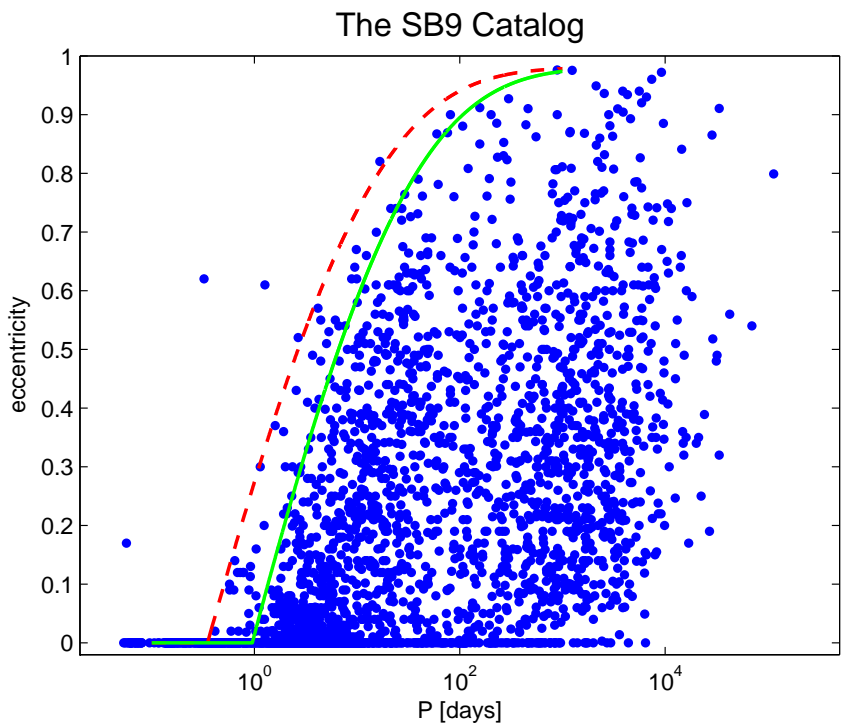}
\caption{The eccentricity of the spectroscopic binaries as a function
  of the orbital period. All 2751 binaries of the SB9 catalog are
  included. The red dashed line is the upper envelope, with an
  equation $f=E-A \exp(-(p B)^c)$, where $E=0.98$, $A=3.25$, $B=6.3$
  and $C=0.23$. The green line is the line under which one can find most
  binaries, with $A=3.5$ and $B=3$}
\label{e_log_p}
\end{figure}

Figure~\ref{e_log_p} shows the eccentricity of the SB9 binaries as a
function of their period. Koch \& Hrivnak (\nocite{koch1981}1981)
already plotted a similar plot, using a much smaller sample that
included only late-type spectroscopic and {\it eclipsing} binaries
known at that time. Similarly to the results of Koch \& Hrivnak,
Figure~\ref{e_log_p} clearly shows that all binaries in the catalog
with periods shorter than 0.35 days, with the exception of one binary,
have circular orbits. In addition, all the binaries with longer
periods except two, have their eccentricities below an 'upper
envelope' that starts at eccentricity zero and climbs up to an
eccentricity of 0.98 when moving from 0.35 to about 350 days. Just
below the upper envelope the space diagram is somewhat sparse, and
only below another stripe does the diagram density become high.

Somewhat arbitrarily, we chose to present the upper envelope and the
sparse stripe by a function of the form

\begin{equation}
f=E-A \exp(-(p B)^c) \ .
\end{equation} 
The values of the parameters of the upper function are $E=0.98$,
$A=3.25$, $B=6.3$ and $C=0.23$, and those of the lower function are
$A=3.5$ and $B=3$.

One could argue that the paucity of eccentric short-period
binaries above the upper envelope is the outcome of binary formation
processes. However, the natural explanation for these two features --- the
circularized binaries and the existence of the envelope --- is that they
are the result of tidal circularization, which is substantially weaker
in long-period binaries. This approach assumes that binaries
were formed above the upper envelope in the diagram, but the
circularization processes pushed them down.

If this is indeed the case, we still have to explain the existence of three
systems high above the upper envelope: the cataclysmic variable WZ Sge,
with a period of $0.057$ days and an eccentricity of $0.17$ (Walker \&
Bell \nocite{walker1980}1980), the binary pulsar PSR 1913+16, with a
period of $0.32$ days and an eccentricity of $0.62$ (Hulse \& Taylor
\nocite{hulse1975}1975) and HD 142315 (Levato \etal\
\nocite{levato1987}1987), a binary with a B8V primary, a period of
$1.264$ days and an eccentricity of $0.61\pm0.2$.

In fact, these three cases might not be a problem for the theory of
circularization:

\begin{itemize}

\item
The eccentricity of WZ Sge might not be real. The derivation of the
radial velocities of WZ Sge was difficult, as the spectra were taken
during the 1978 outburst, and the observed lines probably came from
the outburst material around the white dwarf. In addition, the
velocities derived from the He I lines are consistent with zero
eccentricity (Walker \& Bell 1980).

\item

The binary pulsar is composed of two neutron stars with radii as small
as 1 km (Hulse \& Taylor 1975). Therefore the circularization
processes in that system are not in action at all.

\item

The eccentricity of HD142315 is in fact $0.62\pm 0.20$ (Levato \etal\
1987), so the eccentricity is above the upper envelope by only one
sigma.
\end{itemize}

Figure~\ref{e_log_p} can not be used for a quantitative study of the
circularization processes, especially because of the lack of homogeneity and
the incompleteness of the SB9 catalog. For such a study we will
consider more homogeneous and coeval samples of stars below. Before
diving into the quantitative consideration, we present in the next
subsection two systems that show almost without any doubt an evidence
for long-term circularization. We can acquire such an evidence only for
circular binaries that we have strong reason to believe had 
eccentric orbits in the past.

\subsection{X-ray binaries}

One of the first observational evidences for the long-term
circularization processes taken place in short-period binaries was the
discovery of the circular orbits of the first two close X-ray binaries
--- Cen X-3 (Schreier \etal\ 1972\nocite{sc72}) and Her X-1
(Tananbaum \etal\ 1972\nocite{tan72}), detected by Uhuru, the first
X-ray satellite.  These two X-ray pulsars were found in binaries with
periods of 1.7 and 2.09 days, respectively, and with extremely small
eccentricities. Later analysis of the data of Cen X-3 (Fabbiano \&
Schreier \nocite{fabsc77} 1977) indicated that its eccentricity is
$0.0008 \pm 0.0001$ (see also a review by Bildsten \etal\
\nocite{bildsten1997}1997).  Such precise measurement of the
eccentricity was possible only because of the rapid X-ray pulsation
detected by the satellite, which enabled the observers to follow 
the orbit of the X-ray pulsar in great detail.  The X-ray pulsations come
from a rapidly rotating neutron star (Davidson \& Ostriker
1973\nocite{davidson1973}), formed by a supernova explosion that very
probably produced substantial changes in the binary orbit, inducing a large
orbital eccentricity (Wheeler \etal\ 1974\nocite{wheeler74}; Chevalier
1975\nocite{che75}). Therefore, the observed minute eccentricity had
to be associated with tidal circularization that was in action since
the supernova explosion (Lecar \etal \nocite{lecar76} 1976).

In fact, the short periods of Cen X-3 and Her X-1 might themselves be
another indication for tidal evolution taken place in these systems
during their post-supernova phase. The reason for this is that the
commonly accepted paradigm assumes that the neutron star in these
systems was formed by a supernova type II, which occurred in the last
stages of evolution of the original primary star, when it reached the
supergiant phase (e.g., McCluskey \& Kondo
\nocite{mccluskey1971}1971). The present binary system, with a
separation of order of 0.1 AU, certainly could not accommodate a
supergiant. Therefore, the binary orbits of Cen X-3 and Her X-1 have
probably shrunk because of tidal evolution of the system (e.g.,
Sutantyo 1976\nocite{sutantyo1976}), which led to transfer of angular
momentum from the orbital motion to the present main-sequence star.
Similar ideas were proposed for other X-ray binaries (e.g., Verbunt
\nocite{verbunt1994}1994; van Kerkwijk \etal\ \nocite{van2000}2000;
Janssen \& van Kerkwijk \nocite{janssen2005}2005).

\subsection{Evidence for circularization in early-type
  binaries}
\label{subsection_early_type}

The circularization  occurring in Cen X-3 brings evidence for
tidal interaction in a binary with an early-type star, as the optical
counterpart, V779 Cen (Krzeminski \nocite{krzeminski1974}1974), is of
O6--7 II--III type (Ash \etal\ \nocite{ash1999}1999), with a mass of
about 20 $M_{\odot}$. This is particularly important, as such an early-type
star has a radiative envelope, whereas most of the stars in the SB9
catalog are cool stars with convective envelopes.  Tidal dissipation
is expected to be very efficient in convective envelopes, where the
viscosity is high due to turbulent eddies (Zahn 1975). On the other
hand, in early-type stars with radiative envelopes the dissipative
processes are assumed to be radiative damping on the dynamical tide
(Zahn 1977).  Therefore, the circular orbit of Cen X-3 is an evidence
for tidal interaction of a different kind from that which occurs in
most of the binaries in the SB9 catalog.

The number of spectroscopic binaries with early-type stars is small
because early-type stars have few spectral lines in their spectra,
which makes it difficult to derive their radial velocities and
discover their binarity. Additionally, very hot stars are rare in the
solar neighbourhood, and therefore not many early-type stars are known
as spectroscopic binaries.

In order to study the circularization processes in a larger sample of
early-type stars, a seminal work of Giuricin \etal\
(\nocite{giuricin1984}1984) considered about 200 spectroscopic and
{\it eclipsing} binaries with known O, B and A primaries. They
constructed their sample from the seventh catalog of spectroscopic
binaries (Batten \etal\ \nocite{batten1978}1978) and from the list of
eclipsing binaries of Cester \etal\ (\nocite{cester1979}1979) and Wood
\etal\ (\nocite{wood1980}1980). They plotted the eccentricity as a
function of the orbital period, and showed that the two features of
Koch \& Hrivnak (1981) do appear, although with different
parameters. Excepting only a few cases, all binaries with periods shorter
than $2$ days are circular, and there is an 'upper envelope' that goes
up to an eccentricity of $0.6$.

Giuricin \etal\ (1984) used the fact that the lightcurves of eclipsing
binaries yield one additional parameter --- the fractional radius,
which is {\it not} available for spectroscopic binaries. This
parameter gives the stellar radius as a fraction of the semi-major
axis of the binary orbit. In fact, in early-type stars the
circularization timescale strongly depends on the fractional radius
(see Equation~\ref{circularization_eq}, taken from Zahn
1975). Therefore, Giuricin \etal\ (1984) plotted the eccentricity as a
function of fractional radius for their sample, using a less accurate
spectral estimation for the spectroscopic binaries. They have
discovered that all binaries with fractional radius {\it larger} than
0.3 are circular, and there is an eccentricity 'upper envelope' that
rises from zero to 0.8 when {\it reducing} the fractional radius from
0.3 to 0.05.  Similar results were found in modern photometric data of
early-type binaries in the LMC, as discussed in
Section~\ref{subsection_circ_LMC} (see Figure~\ref{LMC} for a similar
diagram).

The most natural interpretation of these findings is that systems with
fractional primary radius larger than 0.3 have been circularized
during the lifetime of the systems. Note, however, that Giuricin
\etal\ found a few binaries with primary radius larger than 0.3 that
displayed small but significant eccentricities. They attributed these
eccentricities to unseen third distant companions, an effect suggested
by Mazeh \& Shaham (1979; see detailed discussion in
Section~\ref{subsection_MS}).

\subsection{Circularization in coeval sample of binaries}

The first evidence for circularization of coeval binaries was pointed
out by the seminal work of Mayor \& Mermilliod (1984, hereafter MM84).
\nocite{mm84} They have plotted the eccentricities versus period of 33
solar-type main-sequence spectroscopic binaries known at that time in
a few open clusters, including the Hyades, Pleiades, Praesepe and
Coma. MM84 found that the eccentricities of these binaries display a
distinct change at a period of 5.7 days. All seven binaries with
periods shorter than this critical period have circular orbits, while
binaries with periods longer than 5.7 days have orbits with
significant eccentricities.

\nocite{zahn66}  \nocite{zahn75} \nocite{zahn77}
 
The transition period between circular and eccentric binaries can
be directly observed for a given sample of coeval binaries, though
with considerable observational effort. In principle, one can follow
the radial velocities of a large coeval sample of similar stars,
identify the binaries, determine their orbital parameters, the period
and eccentricity in particular, and find the transition period. Since
the seminal work of MM84, a few radial-velocity surveys have detected
the transition-period effect. This was done for binaries with G-type
primaries found in the solar neighbourhood (Duquennoy \& Mayor
\nocite{dm91} 1991), in the open clusters M67 (Latham \etal
\nocite{latham92a}\nocite{latham92b} 1992a,b), NGC 188 (Mathieu \etal
\nocite{mathieu2004} 2004) and recently M35 (Meibom \& Mathieu
\nocite{meibom05} 2005, hereafter MM05), in PMS (Melo \etal\
\nocite{melo01} 2001) and in the halo and field stars (Latham \etal\
\nocite{latham02} 2002). These studies derived the transition periods
for samples with different ages, starting with the PMS binaries, with
an age close to zero, and moving up to the age of the Galactic halo,
which is probably older than 10 Gyrs. The expectation was that
transition periods  lengthen in older samples, for which the
circularization processes have been in action for a longer period of
time. Therefore, the variation of the transition period as a function
of the sample age can be used as indirect evidence of the
circularization processes.

Following this line of reasoning, Mathieu \& Mazeh (1988, hereafter
MM88) \nocite{mm88} even suggested that the cutoff period can be used
to estimate the sample age. However, it turned out that this idea can
not be easily applied, as the theoreticians could not agree upon the
details of the theory of circularization, as discussed below.

\subsubsection{The transition period}
\label{subsub_transition}

Since the early days of circularization studies, the definition of the
observed transition period between the circular and the eccentric
orbits has been unclear. Is the transition period the longest period
for which a circular binary was found, or is it the shortest period
with an eccentric orbit? For example, Figure~2 shows the
eccentricities of 18 binaries with short periods that were found by
the seminal work of MM05 in the open cluster M35, out of their sample
of 32 binaries. We somewhat arbitrarily define as circular binary any
orbit with a derived eccentricity smaller than 0.05. This is indicated
by the horizontal dotted line in the figure. According to this
definition, the longest period with circular orbit is binary no.~4037,
the one with the period of 16.5 days.  On the other hand, the shortest
period with an eccentric orbit is that of binary no.~0422, with 8.2
days (MM05). The two periods, delineated in the figure with vertical
dashed lines, bound the possible range of the transition period. Such
ambiguities exist for most observed samples.

\begin{figure}
\includegraphics[width=13cm]{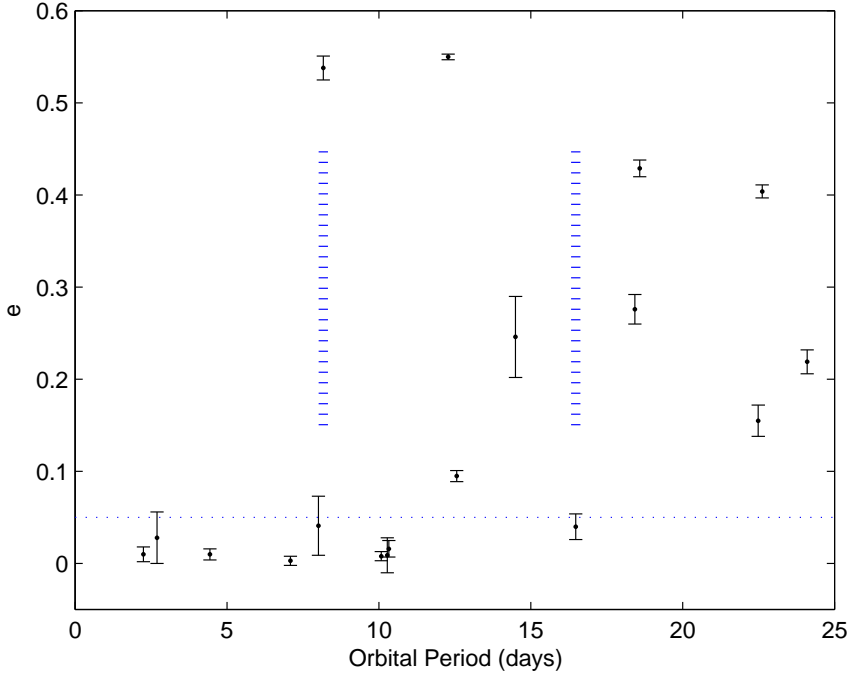}
\caption{The transition range between circular and eccentric binaries
 in M35, after MM05. All binaries with eccentricities below the
 horizontal dotted line are considered circular. The transition range
 is between the shortest period with an eccentric orbit and the
 longest period of circular binary}
\label{range_fig}
\end{figure}

In fact, the width of the transition range is not surprising. Mathieu
and Mazeh (1988) already listed several factors that can blur the
transition period between circular and eccentric binaries, even for a
sample of binaries in which all systems have exactly the same primary
mass. A major factor is the secondary mass, which might be different
for the various binaries in the sample. After all, the gravitational
attraction of the secondary is the source of the tidal force exerted
on the primary, and therefore the tidal circularization timescale of
the binary strongly depends on the binary mass ratio $q=M_2/M_1$,
where $M_1$ is the primary mass and $M_2$ is the secondary mass.  For
a given period and primary star, Mathieu \& Mazeh (1988) got:

\begin{equation}
\tau_{circ}\propto q^{2/3}(1+{1\over{q}})^{5/3} \ .
\end{equation}
Therefore, the secondary mass can lengthen the timescale of the
circularization processes taken place inside the primary by a factor
of 4, when moving from $q=1$ to $q=0.1$. Another factor to take into
account is the circularization processes taking place in the secondary
itself.  For equal masses, the secondary contributes to the
circularization processes as much as the primary. But when we move to
a binary with a mass ratio of, say, 0.5, the circularization processes
in the secondary are negligible, because of the small stellar
radius. Considering the two effects together, the secondary mass can
change the binary circularization timescale by a factor of 8, which
corresponds to a factor of 1.5 in circularization period, if we assume
that $\tau_{circ}\propto P^{16/3}$ (Zahn 1977).

As radial-velocity observations typically are not capable of
determining the nature of the secondary star in single-lined
spectroscopic binaries, the spread of mass ratios in any sample of
spectroscopic binaries induces a spread of the typical circularization
timescales. 

Another factor that can change the time needed for a binary to reach
circularization is its primordial eccentricity. For example, a binary
with initial eccentricity of 0.75 needs twice the amount of time needed by
a binary with initial eccentricity of 0.2 to get to an eccentricity
of 0.05. Therefore, a single circular binary with a long period could
have been the result of a low initial eccentricity, and not
necessarily because of the efficiency of the circularization
processes.

In addition, samples of spectroscopic binaries are not uniform, and
the primary masses are not always the same. A small change in the
primary mass can dramatically change its interaction with the varying
tidal pull of its secondary. 

Lastly, triple systems can show the Mazeh \& Shaham \nocite{ms79}
(1979) effect, by which a third distant star can pump eccentricity
into the binary orbit (see Section~\ref{triple} below), even if the
binary system starts with a circular orbit. Mayor \& Mazeh
\nocite{mm87} (1987) have shown that many of the spectroscopic
binaries might have third faint companions (see a more detailed
discussion in Section~\ref{triple}). Furthermore, Tokovinin \etal\
(\nocite{tokovinin2006}2006; see also Tokovinin
2004\nocite{tokovinin2004}) claimed that more than 90\% of the
short-period binaries, with periods shorter than 3 days, have at least
one additional companion. Therefore, the Mazeh \& Shaham effect can be
quite frequent (see Section~\ref{subsection_MS} for a detailed account
of the observational evidence for this effect), causing (small)
eccentricities to appear in short-period binaries.

Because of all these considerations, the blur of the transition
between circular and eccentric periods in any sample is
inevitable. Therefore, a transit {\it range} between the circular and
the eccentric binary is not that surprising. Nevertheless, some
researchers used the term `cutoff period' to denote the longest period
of a circular orbit in the sample; see, for example, Melo \etal\
(2001) who suggested that the transition period of PMS stars is 7.56
days, based on one circular binary. Such an approach should be
modified, given the different factors that might blur the transition.

\subsubsection{Derivation of the transition period --- 
Meibom \& Mathieu approach} 

A major step in the discussion of the transition period was recently
achieved by Meibom and Mathieu (2005), who came up with a new
algorithm to derive a transition period for a sample of spectroscopic
binaries. Instead of expecting the sample to be drastically divided
between circularized and eccentric binaries, MM05 defined a new
expected `averaged' eccentricity, which presents a smooth transition
between the two parts of the sample. Based on extensive simulations
of populations of binaries, they concluded that the final `averaged'
eccentricity as a function of their period is of the shape

\begin{equation}
e(P)= \left\{
\begin{array}{c c}
0                                         & \mbox{if} \ \  P \le P_{tran}\\
\alpha(1-e^{\beta(P_{tran}-P)})^{\gamma}  & \mbox{if} \  \ P >
P_{tran} \ .\\
\end{array} \right. 
\end{equation} 
In this distribution the eccentricity assumed zero value up to a
'transition' period, $P_{tran}$, and then climbed exponentially to
the averaged eccentricity of $\alpha$. The parameters $\beta$ and
$\gamma$ controlled the steepness of the rise of the averaged
eccentricity. Based on their simulation they adopted the values of
$\beta$ and $\gamma$ to be $0.14$ and $1.0$, respectively. MM05 found
that the mean value of the eccentricity of the long-period binaries in
all samples they considered was $0.35$, and therefore they adopted
this value for $\alpha$.

To derive the transition period of a sample, MM05 suggested finding
the $P_{tran}$ parameter of their function that best fits the sample
periods and eccentricities. The advantage of their approach was that
$P_{tran}$ did not depend on one binary alone, with the longest
circular period or the shortest eccentric period, but, instead took
the whole set of eccentricities and periods in the transition range
into consideration.

In the course of their thorough study, MM05 analysed eight available
samples of coeval spectroscopic binaries with their approach, and
derived transition periods for each of these samples. Their results
are given in Table~\ref{table_circ}, where we round the uncertainties
of the periods.


\begin{table}
\label{tab:circularization}
\caption{Circularization Periods for different coeval samples derived
  by MM05}
  \vskip 1pc
\begin{tabular}{lcr}
Binary Population & $\log Age$ & Transition Period \\
                  &   (Gyrs)     &  (days)\ \, \ \ \ \\
\hline
PMS    & -2.5       &  $7.1\pm 1.2$ \ \ \\
Pleiades            & -1.0 &  $7.2\pm 1.8$ \ \ \\
M35                 & -0.8 & $10.2\pm 1.2$ \ \ \\
Hyades/Praesepe     & -0.2 &  $3.2\pm 1.2$ \ \ \\
M67                 &  0.6 & $12.1\pm 1.2$ \ \ \\
NGC188              &  0.8 & $14.5\pm 1.8$ \ \ \\
Nearby G primaries  & 0.95 & $10.3\pm 2.3$ \ \ \\
Halo & 1.00 & $15.6\pm 2.8$ \ \ \\
\hline
\end{tabular}
\label{table_circ}
\end{table}

\begin{figure}
\includegraphics[width=13cm]{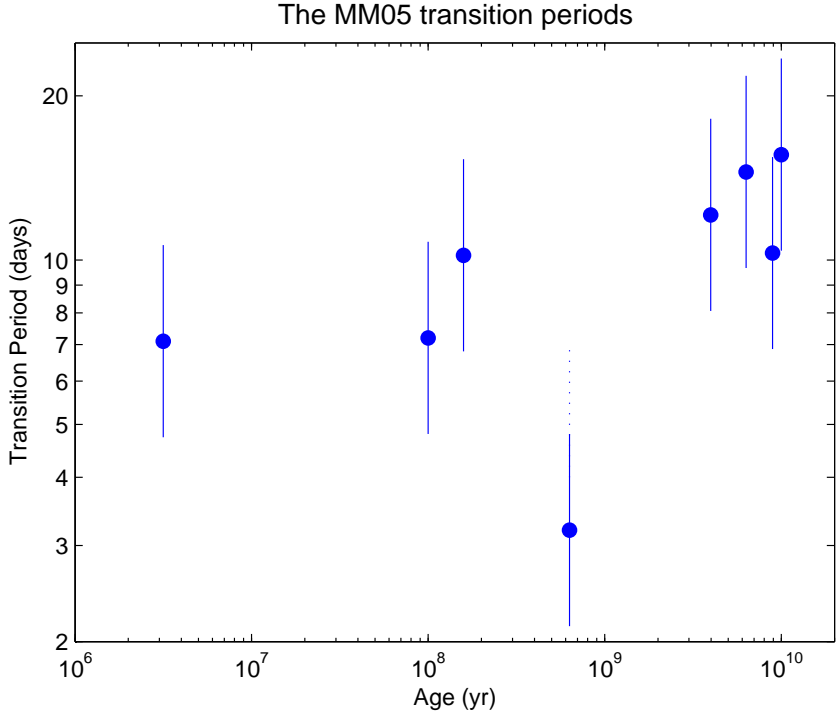}
\caption{The transition period for 8 coeval systems as a function of
their age, after MM05. Contrary to MM05, we assigned to each period an
error of factor of 50\% (see text). The error bar of the transition
period of the Hyades/praesepe sample was extended by a dotted line
(see text).}
\label{circ_per_age}
\end{figure}

Each of the eight samples has similar binaries and is comprised of
main-sequence G-type primaries. It should therefore be interesting
to compare the derived transition period of the samples as a function
of their age. We followed MM05 and plotted their result in
Figure~\ref{circ_per_age}. Based on the discussion above
with regard to the factors that can blur the transition between the
circular and eccentric binaries, we did not use MM05 errors. Instead,
we assign somewhat arbitrarily to each transition period an error of
50\% up and 33\% down, so the up and down errors are the same in
logarithmic scale.

Two features are emerging from the thorough analysis of MM05, as
presented in Table~\ref{table_circ} and
Figure~\ref{circ_per_age}. First, for seven out of the eight coeval
samples discussed by MM05, the transition period is between 6 and 16
days. Second, for the same seven samples, the transition period seems
to be a monotonic increasing function of the sample age. The only
outlier which is not consistent with these two features is the sample
composed of the binaries found in the Hyades and Praesepe, two
clusters of the age of 650 Myrs. The MM05 analysis of this sample
yields a transition period of 3.2 days, but this transition period is
strongly affected by two binaries. One binary is KW 181 (Mermilliod \& Mayor
\nocite{mermi1999}1999), with a period of 5.9 days and an eccentricity
of 0.36, and the other is vB 121, with a period of 5.75 days and an
eccentricity of 0.35 (Griffin \& Gunn
\nocite{griffin1978}1978). Without these two systems, the algorithm of
MM05 would assign a transition period of about 7 days to the Hyades
and Praesepe sample. To mark this fact, we lengthened arbitrarily the
error bar of the symbol that presents the Hyades and Praesepe sample
in the figure.  We therefore conclude that the accumulated
observational data are consistent with the assumption that most, if not
all, transition periods of coeval samples are longer than 6 days, and
there seems to be a tendency for longer transition periods in older
clusters.

\subsection{Confrontation of the theory with the observations}
\label{confrontation}

The analysis of MM05 makes it possible to compare the derived
transition periods with the theory. Three theoretical models were
suggested for the tidal dissipation acting in close binary systems:

\begin{itemize}

\item The theory of equilibrium tide, which assumes that the stellar
  tidal bulge, induced by its companion, lags at some angle after the orbital motion
  of the companion. The lag, which is due to dissipative
  processes inside the star, enables a transfer of energy and angular
  momentum from the orbital motion to the stellar rotation (Zahn
  1966). This mechanism is mostly effective in stars with convective
  envelopes, and therefore the equilibrium tide was applied only to
  binaries with late-type stars (Zahn 1977).

\item The theory of dynamical tides, which assumes that the tidal
  interaction is acting through the damping of gravity modes inside
  the stellar radiative zone (Zahn 1970; 1975; Savonije \& Papaloizou
  \nocite{savpap83} \nocite{savpap84} 1983, 1984). It was assumed that
  this mechanism is only effective in stellar radiative zones, and
  therefore was applied first to high-mass stars with radiative
  envelopes  exclusively (Zahn 1977). However, in the last decade a few studies
  (e.g., Terquem \etal \nocite{ter98} 1998; Goodman \& Dickson
  \nocite{goodman1998}1998; Witte \& Savonije 2002; Savonije \& Witte
  2002) \nocite{witte2002}\nocite{savonije2002} have shown that this
  mechanism can be effective in radiative cores of late-type stars
 as well. Very recently tidal interaction with inertial waves in
  convective envelopes of solar-type stars was also considered
  as a source for orbital evolution (Ogilvie \& Lin 2007; see also
  Ivanov \& Papaloizou 2007). 

\item The theory of hydrodynamic flows, which assumes that the tidal forces
  of the companion induce large-scale meridional flows of the stellar
  surface, which in turn exert torques on the companion (Tassoul
  \nocite{tass95} 1995).  This theory was questioned by Rieutord \&
  Zahn \nocite{rz97} (1997) and was later defended by Tassoul \&
  Tassoul \nocite{tasstass97}(1997).

\end{itemize}

Each of these theories should account for the overall result presented
above which gives {\it an absolute scaling} of the tidal
circularization processes in a binary with G-type primary. The
observationally derived transition periods for the different coeval
samples imply that for a binary with a period of about 10 days the
circularization timescale is of the order of $10^9$ yrs (=Gyr). The
theories should be able to account for the possible mild increase of
the transition period as a function of sample age, presented already
in MM88, and discussed by MM05 in particular.

The plot of MM05 is of great interest because the slope of the period
increase can, in principle, distinguish between the different
theories. This is particularly true if we consider several updated
versions of the equilibrium tide theory. This theory was restudied by
Zahn (\nocite{zahn1989}1989), who specifically considered the reduction
of the turbulent viscosity when the tidal period becomes shorter than
the convective turnover time. He concluded that the circularization
timescale should be $\tau_{circ}\propto P^{16/3}$. In a follow-up paper,
Zahn and Bouchet (1989) concluded that all the circularization of
G-type binaries must occur during the pre-main-sequence (=PMS)
phase, when the stars are substantially larger than they will be during their
main-sequence phase. They further claimed that the circularization during
the main-sequence phase is negligible, and they therefore predicted that the
circularization period would be between 7.2 and 8.5 days for all
samples, regardless of their respective ages.

Not all studies agreed with Zahn (1989) approach to the reduction of
the turbulent viscosity. Goldman \& Mazeh (\nocite{goldman1991}1991),
for example, proposed a modified approach that resulted in
$\tau_{circ}\propto P^{10/3}$. They also pointed out the evidence
that was available at that time for the variation of the transition
period with the sample age. Goldreich \& Nicholson
(\nocite{goldreich1989}1989) and Goodman \& Oh (\nocite{goodoh97}1997)
studies yielded power dependence which is in between that of Goldman
\& Mazeh and that of Zahn (1989).

It seems that the plot of MM05 (Figure~\ref{circ_per_age}) indicates
that the circularization processes during the PMS phase do circularize
binaries up to about 6--8 days, as Zahn and Bouchet suggested. It also
suggests that transition periods probably become longer in older
samples, in contrast to the other claim of Zahn \& Bouchet (Goldman \&
Mazeh 1991; Mathieu \etal\ \nocite{mathieu1992}1992), although the
slope of the figure is not conclusive enough to distinguish between
the different theories. If this is true, we must find out why the
theory of equilibrium tide, when applied to {\it main-sequence} stars,
yields a circularization absolute efficiency which is too low to
explain the observations (e.g., Goodman \& Oh 1997; Terquem \etal\
1998; Goldman \& Mazeh 1991; Sasselov \nocite{sasselov2003}2003).  We
therefore have to close this discussion with the banal statement that
we need more data for a definite conclusion.

\subsection{Circularization of binaries with giant star component}
\label{giant}

An additional line of evidence for tidal circularization was studied
by Verbunt \& Phinney (\nocite{verbunt1995}1995, hereafter VP95), who
considered the eccentricity as a function of orbital period of
binaries containing a giant component found in open clusters. VP95
noted that giants can serve as the simplest test cases for the
circularization theory, because the orbital period of such systems is
much larger than the eddy timescale, and therefore we do not need to
worry about the reduction efficiency of the viscosity. Furthermore,
giants are fully convective and therefore avoid the problem of tides
in the radiative zone. We note that the large radii of the giants
render the circularization during PMS phase irrelevant, another
advantage of studying binaries with giants. It is true that the giants
have an history of fast expansion, which causes the circularization
timescale to change rapidly. However, VP95 claimed that from the
cluster age and turnoff mass one can reconstruct the history of the
giants and integrate the expected theoretical circularization effect.

The sample of VP95 included 28 binaries from 12 open clusters, taken
from the works of Mermilliod \& Mayor (\nocite{mm89}\nocite{mm90}1989;
1990), Mermilliod \etal\ (\nocite{mermilliod1989}1989) and Mathieu
\etal\ (\nocite{mathieu1990}1990). They found that except two systems,
all binaries with periods shorter than 200 days have circular orbits.
However, as the systems are with different ages and masses, the
orbital period is not the best parameter to compare the observed
eccentricity with. When VP95 considered instead the {\it expected}
eccentricity reduction, based on the history of the giants derived
from their mass and position on their H-R diagrams, they concluded
that all are consistent with Zahn theory.

More than a decade has passed since the seminal work of VP95, and many
more giants have been discovered since. A new study of the binaries
with giants in open clusters and in the field is highly due, so we can
learn more about the circularization processes and derive their
absolute scaling of efficiency.

\subsection{Highly eccentric binaries and the theory of circularization}

Some of the main-sequence spectroscopic binaries were found to have
very high eccentricities. The five most eccentric binaries in SB9,
with an eccentricity larger or equal to 0.95, are compiled in Table 2,
which lists their orbital periods and eccentricities. The two most
eccentric systems might be a challenge to the circularization theory,
as their periastron distance is quite small, and therefore we would
expect them to circularize, or at least to substantially reduce their
eccentricities on a relatively short timescale (Duquennoy \etal\
1992).
\begin{table}
\label{tab:eccentric}
\caption{The five most eccentric spectroscopic binaries in SB9. The
table includes the period and eccentricity of each system and also the
`periastron period', $P_{peri-dist}$ --- the orbital period of a
circular orbit with a radius that equals the actual periastron
distance.}
\vskip 1pc
\begin{tabular}{lrccl}
Binary      & $e$ \ \ \ & Orbital Period & $P_{peri-dist}$&Reference\\
            &            &  (days)       &  (days)\\
\hline
HD 2909   &      0.949   &  2128   & 25   & Mazeh \etal\ 1995   \\
          & $\pm$0.002   &                             &\\
HD 165590 &      0.958   &  7400   & 64   & Batten \etal\ 1979  \\
          & $\pm$0.001    &                            &\\
HD 123949 &      0.972  & 9200    & 43 & Udry \etal\ 1998 \\
          & $\pm$0.057            &                                &\\
Gliese 586A  &      0.9752 &  890 & 3.5 & Duquennoy \etal\ 1992 \\
          & $\pm$0.0003 &  \\
41 Dra    &      0.9754 & 1250 & 4.8  &  Tokovinin \etal\ 2003   \\
          & $\pm$0.0001\\
\hline
\end{tabular}
\end{table}
\nocite{tokovinin2003}
\nocite{duq1992}
\nocite{griffin1984}

\nocite{udry1998} \nocite{batten1979} \nocite{mazeh1995}

To estimate the effectiveness of the tidal forces at periastron, I
introduce the `periastron period', $P_{peri-dist}$ --- the orbital
period of a circular orbit that has the same orbital radius as
the actual periastron distance. It is easy to show that
\begin{equation}
P_{peri-dist}=P(1-e)^{3/2} \ . 
\end{equation}
Note that this is {\it not} the period that presents the orbital
angular velocity at periastron. Table 2 lists the corresponding
periastron periods of the five systems.

>From the table we see that two of the systems, with eccentricities
larger than 0.97, indeed have short periastron distances, which are
equivalent to orbital periods shorter than 5 days. If these systems
are of the age of a few Gyrs, we would naively expect them to
circularize, or at least to reduce their eccentricities during their
main-sequence lifetime.

Goldman \& Mazeh (\nocite{goldman1994}1994), who studied the tidal
history of Gliese 586A, argued that for highly eccentric binaries the
tidal shear, which drives the turbulent viscosity, changes near
periastron on a timescale shorter than the convective
timescale. Therefore the turbulent viscosity (e.g., Hut 1982a;b)
\nocite{hut1982a}\nocite{hut1982b} is substantially reduced, as was
argued in Section~\ref{confrontation}. With their recipe for 
the reduction of turbulence, Goldman \& Mazeh were able to suggest a tidal
history of Gliese 586A that leads to the present parameters of the
system. Obviously, similar arguments can be applied to the other
systems.

Goldman \& Mazeh even argued that the highly eccentric binaries have
the potential for distinguishing between the different recipes for
viscosity reduction. Unfortunately, very eccentric binaries are
difficult to discover, as most of their radial-velocity variation is
concentrated in a very small part of the orbital period (e.g., Griffin
1984). Therefore only very few highly eccentric binaries are
known. Nevertheless, I suspect that many of these binaries exist, and
when they are discovered we will be able to use them as test cases for
tidal interaction.

%
\section{Synchronization}    %
\label{section_synchro}
%

As stated in the introduction, tidal forces in binaries also tend to
synchronize the rotation of the two stars with the orbital motion. The
theory of tidal evolution (Zahn \& Bouchet \nocite{zahnbou89}1989;
Witte \& Savonije 2002) predicts that the synchronization timescale
will be two or three orders of magnitudes shorter than the
circularization timescale. This has to do with the fact that the
angular momentum associated with stellar rotation is much smaller than
the orbital angular momentum. Consequently, the angular momentum
transferred between the orbital motion and the stellar rotation in
order to reach synchronization is much smaller than that which is  needed to
attain circularization. This means that we can expect binaries to
reach synchronization a long time before they  reach
circularization. Similarly to circularization, we should also expect
coeval sample of binaries to show a transition period between
synchronized and non-synchronized binaries. Such an observed
transition can serve as another confirmation of the theory of tidal
interaction.

However, both the observations and the theory of stellar
synchronization are more complicated than those of tidal
circularization. Even the term `stellar rotation' is not well defined,
as the notion of a star rotating as a rigid body is obviously an
oversimplification. For example, differential rotation as a function
of stellar latitude (e.g., Lyytinen \etal\ \nocite{lyytinen2002}2002;
Croll \etal\ \nocite{croll2006}2006; Walker \etal\ \nocite{walker2007}
2007) must be taken into account when we define stellar
rotation. Therefore the following section adopts an observational
approach, according to which the stellar rotation is the observed
rotation, and which disregards complications induced by differential
rotation.

The stellar rotational period can be obtained either by photometric or
spectroscopic observations. Photometric monitoring can reveal stellar
{\it periodic} variability, which presumably is caused by stellar spots
that come into and go from the facing hemisphere of the star,
as the star rotates on its axis (e.g., Stauffer \etal\
\nocite{stauffer1987}1987). Stellar spectra, with high enough
resolution and S/N ratio, yield line broadening which can be modelled
to yield stellar rotational velocity. With the knowledge of the size
of the stellar radius and inclination angle it is possible to derive
the stellar rotation period (e.g., Stauffer \etal\
\nocite{stauffer1984}1984). However, as stellar radii and inclinations
are usually not well known, the spectroscopic approach in most cases
induces a large uncertainty of the rotational period. On the other hand, in
spectroscopic binaries the spectra have already been obtained, in
contrast to the photometry that usually necessitates additional
observational effort.

The variability method is dependent upon the longevity of the spots.
We assume that although stellar spots indeed form and dissolve, they
still maintain some coherence with a timescale much longer than the
stellar rotation period. If this is the case, then the spots can
maintain a periodic brightness variability, modulated with the
rotation period. However, as stellar spots form and dissolve, the
amplitude and phase of the modulation might not be constant (see, for
example, a discussion of the stellar spots of V1794 Cyg by Jetsu
\etal\ \nocite{jetsu1999}1999). This can cause some spread of the
frequency power of the modulation over neighbouring frequencies,
rendering the detection of the periodic modulation more
difficult. Therefore many measurements are needed in order to detect
the rotational frequency, preferably distributed over many rotational
periods.

Another complication faced by the photometric approach is the
existence of other sources of variability, coming either from the star
itself, like stellar pulsation, or from the photometric modulation of
its companion. The latter is specifically true for binaries with
comparable brightness, where the secondary modulation can have an
amplitude that is comparable with that of the primary. In addition, the
ellipsoidal variability, with half the binary period, can also be
present in the photometric data (see, for example, the analysis of the
RS CVn-type stars RT And by Zhang \& Gu \nocite{zhang2007}2007 and SZ
Psc by Eaton \& Henry \nocite{eaton2007}2007). Therefore, the
identification of stellar rotation by a periodic modulation
should be done with extra caution, and only after other sources of
periodic modulation have been excluded.
 
On the theoretical side, whereas for the eccentricity evolution only
the tidal interaction is in action, stellar rotation has its own
evolution even in single stars. Contraction of PMS stars that are
approaching the main-sequence phase and their interaction with their
accretion disc (e.g., K\"onigl \nocite{konigl1991}1991; Shu \etal\
\nocite{shu1994}1994) can change the stellar rotation substantially
(for observational evidence see, for example, Choi \& Herbst
\nocite{choi1996}1996 and Stassun \etal\ \nocite{sta99}1999). In
addition, the rotational angular momentum itself can evolve during the
main-sequence phase (e.g., Baliunas \etal\ \nocite{baliunas1995}1995), via
magnetic coupling with the stellar wind, for example.

Because of the difficulties of the theory and observation, less
attention has been paid by the binary community to the study of tidal
synchronization. Two exceptions are the works of Abt \etal\
(\nocite{abt2002}2002) and Abt \& Boonyarak (\nocite{abt2004}2004),
who studied samples of B and A stars known to reside in binaries. Abt
\etal\ (2002) derived the rotational period of their samples from the
stellar line broadening, and concluded that all B-type stars with
orbital periods shorter than 2.4 days are synchronized, while binaries
with periods between 2.4 days and 5.0 days are `nearly
synchronized'. The corresponding period ranges for A-type stars are
4.9 and 10.5 days, or twice as large. They also found that the
rotations of the primaries are synchronized earlier than their orbits
are circularized. The maximum orbital period for circularized B
binaries is 1.5 days and for A binaries is 2.5 days. Abt \etal\
suggested that their finding is consistent with the theory of Zahn for
early-type stars. 

Abt \& Boonyarak found that B and A stars in binaries with
periods as long as 500 days have rotational period significantly
shorter than the corresponding single stars. They therefore concluded
that the synchronization processes can have impact on the stellar
rotation even for relatively wide binaries with periods as long as 500
days. The origin of these findings is still not clear, and needs
further theoretical study.

The results of these two studies are most interesting, despite the
difficulties to derive the rotational stellar periods, specially
because these two studies focus on early-type stars (see
Subsection~\ref{subsection_early_type} for discussion of the very few
studies of circularization of early-type stars). We wish that similar
studies would have been performed for G and K stars, and the results
confronted with the theory. We could benefit from enlarging the
interaction between theory and observations to include both
circularization and synchronization of both early- and late-type stars. 

Note that because the synchronization timescale is probably
substantially shor\-ter than that of circularization, the synchronized
binaries might be more common than the circularized orbits (see
above). This means that we can find many synchronized binaries with
eccentric orbits.  However, for eccentric binaries the orbital angular
velocity is not constant, and therefore the synchronization frequency
is not well defined. There is no rotating frame of reference attached
to the primary for which the secondary is at rest. Therefore,
eccentric binaries need a new definition of synchronization which will
be reviewed in the next subsection.

\subsection{Pseudo-synchronization}

As the tidal forces strongly depend on the distance between the two
stars, it is clear that in eccentric binaries each star should
respond mostly to the orbital angular velocity near the periastron,
where its companion is the closest. In a seminal paper, Hut
(\nocite{hut81}1981) has demonstrated that in the so-called weak friction approximation
(see Zahn in this volume),
each star achieves its equilibrium when it rotates with a frequency that is 
smaller but still close to the orbital periastron frequency. The equilibrium
frequency, which Hut calls the pseudo-synchronization frequency,
is
\begin{equation}
n_{pseudo}=\frac{1+\frac{15}{2}e^2+\frac{45}{8}e^4+\frac{5}{16}e^6}
{(1+3e^2+\frac{3}{8}e^4)(1-e^2)^{3/2}}\,n_{orbit} \ ,
\end{equation}
where $n_{orbit}=2\pi/P$ is the {\it averaged} orbital
frequency. Sometimes it is convenient to express the
pseudo-synchronization frequency with the angular velocity at
periastron --- $n_{peri}$:
\begin{equation}
n_{peri}=\frac{(1+e)^{1/2}}{(1-e)^{3/2}}\,n_{orbit} \ ,
\end{equation}
which leads to the expression
\begin{equation}
n_{pseudo}=\frac{1+\frac{15}{2}e^2+\frac{45}{8}e^4+\frac{5}{16}e^6}
{(1+3e^2+\frac{3}{8}e^4)(1+e)^{2}}n_{peri} \ 
\end{equation}
(Hut 1981).
To third order of the eccentricity $e$ one gets:
\begin{equation}
n_{pseudo}\simeq
(1-2e+7.5e^2-13e^3)\,n_{peri}=
(1+6 e^2)\,n_{orbit} 
\ .
\end{equation}

Only subsequent to the work of Hut (1981) can we examine the eccentric
orbit for its pseudo-synchronization. Of particular interest are
coeval samples of binaries with a transition between the
pseudo-synchronized and non-pseudo-synchronized binaries.

\subsection{Synchronization of young stars}

Two important studies of synchronization of short-period binaries
have been performed recently. Marilli \etal\ (\nocite{mar07}2007)
photometrically observed 40 PMS T Tauri stars that were discovered as
X-ray sources by the ROSAT Survey of the Orion complex, and
subsequently were observed optically with both low and high resolution
spectroscopy. They obtained extensive photometric data and analysed
them with the Lomb-Scargle periodogram (Lomb \nocite{lomb76}1976;
Scargle \nocite{scar82}1982), deriving photometric modulation periods
for 39 stars. Assuming the modulations are due to spots, these periods
present the stellar rotational periods. Meibom \etal\
(\nocite{mei06}2006) derived the rotation periods of 13 close binaries
in M34 and M35, two young open clusters, with an age of 150 and 250
Myrs, respectively.


\begin{table}
\label{tab:pseudo}
\caption{The PMS and M35 young cluster eccentric systems with their
  rotation periods from Marilli \etal\ (in the first part of the
  table) and Meibom \etal\ (below the horizontal line in the
  table). For each system the pseudo-synchronized period is calculated
  according to the Hut (1981) recipe. The last system is the only circular   system with a non-synchronized rotational period from M35.}
\vskip 1pc
\begin{tabular}{lcccc}
Binary & Orbital & e\ \ & Pseudo-synch. &Rotational \\
       &  Period &     &    Period     & Period    \\
                  &   (days)     & & (days)&  (days)\\
\hline
044059.2-084005    & 13.56     &  0.22 & 10.5   &  2.75 \\
053043.1-043453    & 40.57     &  0.32 & 24.7   & 12.90 \\
053202.1-073153    & 46.85     &  0.47 & 18.4   &  3.44 \\
0539.8-0205        &18.74      &  0.39 &  9.4   &  4.48 \\
\hline
06091557+2410422   & 8.17      &0.65   &  1.6   &  3.71 \\
06095563+2417454   &30.13      &0.27   &  20.8  &  2.84 \\
06085441+2403081   & 12.28     & 0.55  &  3.7   &  6.03 \\
06090257+2420447   & 10.28    &  0.001 & 10.3   &   2.3
\end{tabular}
\label{tab_synchro}
\end{table}

Eight of the stars observed by Marrilli \etal\ were found to be in
spectroscopic binaries (Covino \etal\ \nocite{covino2001}2001).  Four
systems, with orbital periods shorter than 10 days, were found to be
circular and synchronized. The other four binaries, with orbital
periods between 10 and 50 days, with eccentric orbits, are listed
in the upper part of Table~\ref{tab_synchro}. The table shows the
orbital, rotational and pseudo-synchronization periods.  It is easily observed that the rotational periods are substantially shorter than the
pseudo-synchronization periods.

\begin{figure}
\includegraphics[width=11cm]{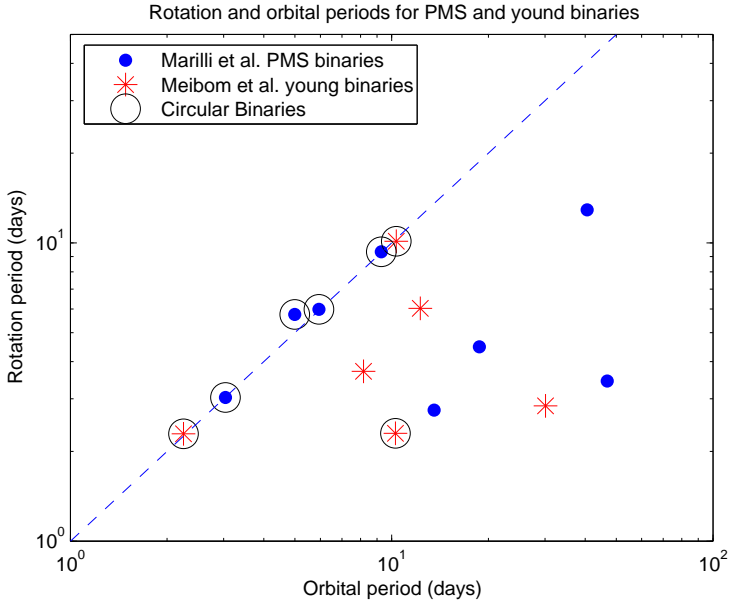}
\begin{centering}
\caption{Rotational periods as a function of the orbital periods for
eight PMS binaries (Marilli \etal\ 2007) and six young stars (Meibom
\etal\ 2006). Circular binaries are encompassed by a circle. The
dashed line is the locus of points for which the rotation period
equals to the orbital period.}
\label{fig_synchro}
\end{centering}
\end{figure}

Six systems from the Meibom \etal\ (2006) study have periods shorter
than 100 days, and of the six, three systems have eccentric
orbits. These systems are included in Table~\ref{tab_synchro},
together with the system 06090257+2420447, for which the orbit is
circular but is very far from synchronization. All binaries in the two
studies with periods shorter than 100 days, including those which are
documented in the table and another six circular {\it and}
synchronized binaries, are plotted in Figure~\ref{fig_synchro}.

Both samples are extremely small, but one feature seems to emerge from
the figure: the transition period between circular and eccentric
orbits is between 8 and 10 days. It would be reasonable to expect that
the transition between pseudo-synchronized and non-pseudo-synchronized
binaries should appear on {\it much longer} period, as the
synchronization processes are a few orders of magnitude more
efficient. However, the striking feature of this figure is that the
transition between pseudo-synchronized and non-pseudo-synchronized
systems is at about the same period!

To understand this surprising similarity between the two transition
periods we must  bear  in mind that while the circularization
processes worked undisturbed during the PMS phase with high efficiency
(Zahn \& Bouchet 1989), the synchronization processes were drastically
affected by the internal variation inside the stars which changed their
radii and therefore their rotational period. The first 100 Myrs of M35
after the stars settled on the main sequence were probably not enough
to achieve pseudo-synchronization for binaries with periods
longer than 8-10 days. In order to observe the strong
potential of the synchronization processes, we apparently need to
observe much older clusters.

Note that for the two most eccentric systems in Table 3, with
eccentricity of 0.55 and 0.65, the pseudo-synchronized period is {\it
shorter} than the actual rotational period of the primary star.  This
is in contrast to the other five systems with more moderate
eccentricities, for which the pseudo-synchronized period is {\it
longer} than the rotational period. If this feature is validated
by further observations, the difference between the highly eccentric
binaries and the binaries with small or medium eccentricities could be a
noteworthy feature (see, however, the discussion of V1143 Cyg in
Section~\ref{V1143}. The observed longer rotational periods could be
associated with the approximation used by Hut in his work, which did
not take into account the reduction of the turbulent viscosity on a
timescale shorter than the convective timescale (Goldman \& Mazeh
1994). It is possible that the pseudo-synchronization periods for some stars in
highly eccentric orbits are longer than the ones derived by Hut's
formula. It is therefore extremely important to derive stellar
rotational periods for many more highly eccentric binaries and to compare
them with the pseudo-synchronized periods.

\subsection{Interesting cases}

\subsubsection{V1143 Cyg}
\label{V1143}

V1143 Cyg is a bright, nearby, eclipsing double-lined spectroscopic
binary, and therefore allows, in principle, for a determination of the
masses, radii and luminosities of the two stars (e.g., Popper
\nocite{popper1980}1980). The binary consists of two F5V stars
orbiting each other with a period of 7.6 days and an eccentricity of
0.54 (Snowden \& Koch \nocite{snowden1969}1969). Apparently, at the
age of 2 Gyrs, the system did not have enough time to circularize (Andersen
\etal\ \nocite{andersen1987}1987). Andersen \etal\ derived the
rotational broadening of the two components to be $V_A \sin i_A =
18\pm3$ and $V_B \sin i_B = 28\pm3$ for the primary and secondary,
respectively. With the derivation of the two radii --- $1.346\pm0.023$
and $1.323\pm0.00023$ solar radii for the primary and secondary, and
assuming an inclination close to $90^{\circ}$, one gets for the rotation
period of the two stars 3.8 and 2.3 days, respectively.  Andersen
\etal\ results are confirmed by the very recent detailed study of
Albrecht \etal\ (2007).

Andersen \etal\ (1987) pointed out that the pseudo-synchronization
period, derived from the orbital period and the eccentricity, is 2.3
days, exactly equal to the rotation period of the
secondary. Therefore, V1143 Cyg is a good example of a binary that has
reached pseudo-synchronization, at least for one star, but obviously
has  not yet reached circularization. Thus, V1143 Cyg could be
considered  one of the few binaries that strongly supports Hut's
(1981) theory of pseudo-synchronization. 

However, the rotational period of the primary is a challenge to the
theory. It is hard to conceive how  the secondary could have been
pseudo-synchronized while the primary was not, especially since 
the finding of Andersen \etal\ (1987) indicates that the primary and the
secondary are extremely similar. Is it possible that the initial rotation of
the primary and the secondary were so different that there was enough
time to pseudo-synchronize the secondary but not the primary? I find
such an explanation hard to believe. If it is indeed true, the disparity
between the rotational periods of the two components remains a
mystery. The primary of V1143 Cyg joins the group of highly eccentric
binaries with rotational periods longer than Hut's
pseudo-synchronization period, which was discussed above. 

Recent work on the inclination of the two stars of V1143 Cyg
relative to the orbital plane is reviewed in Section~\ref{V1143_RM}.

\subsubsection{KH 15D}
\label{KH_15D}

KH 15D is a T Tau binary star in the young cluster NGC 2264 (Badalian
\& Erastova \nocite{badalian1970}1970) that fades by three mag every
48 days (Kearns \& Herbst \nocite{kearns1998}1998). The best model to
account for the many observed details of the periodicity of KH 15D
assumes that the binary, with an orbital period of 48 days, is
surrounded by a precessing circumbinary disk (e.g., Herbst \etal\
\nocite{herbst2002}2002). Because of the binary motion, the primary
and the secondary are being occulted periodically by the disk (e.g.,
Hamilton \etal\ \nocite{hamilton2003}2003; Chiang \& Murray-Clay
\nocite{chiang2004}2004; Winn \etal\ \nocite{winn2004}2004; 2006).

Hamilton \etal\ (\nocite{hamil05}2005) measured a rotational period of
9.6 days for the primary star of KH 15D, based on photometric
periodicity observed out of eclipse. {\it Assuming}
pseudo-synchronization, Herbst \& Moran (\nocite{herbst2005}2005) and
Hamilton \etal\ (2005) derived an eccentricity of
$e=0.65\pm0.01$. This was probably consistent with the eccentricity of
$0.68 \leq e \leq 0.8$, derived by Johnson \etal\
(\nocite{johnson2004}2004) from Keck radial-velocity
measurements. Indeed, Winn \etal\ (2006) derived a value of
$e=0.574\pm0.017$ for the orbit of KH 15D in one of their models. If
this value is correct, then the primary star in KH 15D system rotates
with a frequency close to its pseudo-synchronization frequency,  despite the long orbital period of 48 days. 
Additional spectroscopic and photometric observations are certainly needed in
order to establish the rotational period of the primary, the
eccentricity, and then the pseudo-synchronization of this fascinating
system.

Recent work on the stellar inclination of KH 15D relative to its
orbital motion is reviewed in Section~\ref{KH_RM}.

\section{Alignment}       %
\label{section_alignment} %

As stated in the introduction, tidal forces in binaries tend also to
align the rotation axes of the two stars with the orbital angular
momentum. As argued for synchronization, the alignment timescale
should be two or three orders of magnitudes shorter than the
circularization timescale. This is so because the angular momentum
transferred between the orbital motion and the stellar rotation in
order to reach alignment is much smaller than that which is needed to attain
circularization. This means that we can expect binaries to reach
alignment long before reaching circularization, and we expect all
circularized binaries to be aligned.

Contrary to circularization and synchronization, we do not know if
binaries are formed with misaligned rotation. On the contrary, one
could argue that binary systems are formed with stellar spin aligned
with the binary motion. This is so because both stellar rotation and
orbital motion are relics of the same angular momentum of the
primordial cloud, out of which the binary was formed.  Common
primordial spin-orbit misalignment can rule out some binary formation
scenarios (e.g., Durisen \etal\ \nocite{durisen2001}2001; Bate et
al. \nocite{bate2002}2002; Bonnell \& Bate \nocite{bonnell2005}2005;
Machida 2005). \nocite{machida2005} Thus, we do not know if tidal
interaction to attain alignment should set into action at all. It is
therefore important for our understanding of binary formation to
measure the present extent of the spin-orbit inclinations in the
binary population.

Observationally, the spin-orbit inclination, $i_{rel}$, is 
difficult to obtain, and consequently not much work has been done on binary
alignment.  One way to derive the relative inclination is to use
stellar spectra with high enough resolution and S/N ratio that can
yield detailed line profiles. The line profiles can be modelled to give
projected stellar rotational broadening --- $V_*\sin i_*$, where
$i_*$ is the inclination of the stellar rotation axis relative to
our line of sight. With an {\it independent} knowledge of the size of
the stellar radius {\it and} rotational period one can estimate the
inclination angle $i_*$.

Note that this approach allows us to measure the stellar spin inclination
relative to our line of sight and {\it not} relative to the binary
orbit. In order to derive $i_{rel}$ we also need to know $i_{binary}$
--- the orbital inclination of the binary. This angle is not known for
spectroscopic binaries, and can be derived directly from the
observations only for eclipsing binaries. The difference between the
two inclinations, $i_{binary}-i_*$, is an estimate of the {\it
minimum} relative angle $i_{rel}$.

In addition, we do not know the stellar radius and the
rotational period  for most stars. The stellar radii can be directly obtained only
from observations of eclipsing binaries. Estimation of the
stellar radii from the spectral type is not very
accurate. The stellar rotational period is also not well established, and
can be obtained either via special photometric monitoring of the system
(see above), or by {\it assuming} synchronization.  As a result of 
these drawbacks, the line broadening approach to studying binary
alignment was not used extensively in the past. In fact, in some cases
the derivation is reversed, and is used to estimate the rotational
period (see Section~\ref{section_synchro}). Another example is the work of Beatty
\etal\ (\nocite{beatty2007}2007), which derived the radius of the
primary for the eclipsing single-lined binary HAT-TR-205-013 from the
observed $V\sin i_*$, assuming synchronization and alignment.

As discussed in Section~\ref{DI_Her}, spin-orbit misalignment can
cause three noteworthy effects. First, it can change the apsidal
precession of the binary by introducing counter apsidal precession. As
in the case of DI Her, such an effect can, in principle, be measured
if the actual precession is observed and the other precessions can be
derived. Second, the binary tidal interaction will cause the stellar
spins {\it and} the orbital plane to precess around the total angular
momentum of the system with the same period. Such a forced precession
will change the stellar spin relative to our line of sight, resulting
in a change of the stellar rotational broadening of the two stars.
Third, it might also slightly change the inclination angle of the
binary, which can be detected by radial-velocity measurements, as well
as by photometry if the binary is an eclipsing system. However, a
study of the old observations of DI Her by Reisenberg \& Guinan (1989)
is one of the few works that tried to use this indirect line of
evidence for detecting spin-orbit misalignment (see, however, Mayor \&
Mazeh (1987) for a different misalignment found in triple systems and
the observational search for its effects).

Evidence for significant spin-orbit misalignment should have
implications for our understanding of other systems. For example, a
forced precession induced by spin-orbit misalignment was already
considered in the context of X-ray binaries. It was suggested by
Roberts (\nocite{roberts1974}1974) to account for the 35-day
periodicity of Her X-1 (Giacconi \etal\ \nocite{giacconi1973}1973), and for
the 164-day precession of the relativistic jets of SS 433 (see Margon
\nocite{margon1984}1984). This idea was never confirmed for Her X-1,
and a different precession model of Katz (\nocite{katz1973}1973)
became more popular. For SS 433 the evidence is also not conclusive
(see a review of SS433 recent observations and theory by Fabrika
\nocite{fabrika2004}2004). Therefore, our comprehension of the typical
spin-orbit inclination in binaries is of great importance. It is a
pity that the derivation of this inclination from indirect
observations is so difficult.

Another, more direct, method to derive the spin-orbit relative
inclination angle is through the Rossiter-McLaughlin effect, which is
relevant only for eclipsing binaries. This effect is described in
detail in the next section.

\subsection{The Rossiter-McLaughlin effect}
\label{rm_effect}

The Rossiter-McLaughlin (RM) effect is named after Rossiter
(\nocite{rossiter1924}1924) and McLaughlin
(\nocite{mclaughlin1924}1924) who observed the effect for $\beta$
Lyrae and Algol --- two famous eclipsing binaries, within the same year.

The RM effect is the varying shape of the stellar line profile during
eclipse and is based on a combination of two very basic ideas. The
first is stellar rotational broadening, which is due to the fact that different
sections of the stellar disk, at different distances from the
projected stellar rotational axis, have different Doppler shifts
(e.g., Gray \nocite{gray2005}2005). The second is that during the
progress of an eclipse, varying parts of the stellar disk of the
eclipsed component are covered by the eclipsing star. Therefore, the
line shape of the eclipsed star is bound to vary during the
eclipse. In fact, the eclipsing body allows us, through observing the
RM effect, to indirectly resolve the disk of the eclipsed star. Such an
effect was observed, for example, by Rauch \& Werner
(\nocite{rauch2003}2003) in their observations of AA Dor.

The RM effect has one additional interesting feature. If the orbital
motion of the companion is in the same direction as the stellar
rotation, then during the first part of the eclipse the blue-shifted
part of the stellar disk is eclipsed, while the red-shifted part of the stellar disk is
eclipsed during the second part of the eclipse. If, on the other hand,
the orbital motion and the stellar rotation are in opposite
directions, then the red-shift part of the disk is eclipsed
first. Thus, the RM effect can also detect the relative direction of
stellar rotation.

Dividing the eclipse into two periods --- one when the red-shifted
side of the disc is eclipsed, and the other when the blue-shifted side
is eclipsed, might be illuminating on a conceptual level. For example,
if the stellar axis is aligned with the orbital motion, then we expect
these two parts of the eclipse to be symmetrical for circular
orbits. Asymmetry between the two parts clearly indicates a nonzero
relative inclination.

The RM effect recently attracted new interest because of its
application to extrasolar planets (e.g., Queloz \etal\
\nocite{queloz2000}2000; Winn \etal\ \nocite{winn2007}2007; Loeillet
\etal\ \nocite{loeillet2007}2007).  However, contrary to the RM effect
on a transiting planet, in a binary system the derivation of the
relative inclination is not easy, specifically due to the
contribution of the secondary star to the total light of the
system. In order to follow the distortion of the primary lines when
the primary is eclipsed, for example, one  must resolve the blending
of the primary and secondary lines. This might be complicated whenever the
secondary brightness is comparable to the primary brightness, because
during the eclipse there is almost no shift between the velocities of
the two stars, except for some very eccentric orbits. Such a problem
does not exist when analysing the RM effect of transiting planets, as
the planet's light is negligible. On the other hand, the amplitude of
the effect in a binary is much larger than in that of a planet's
transit, as a planet covers during its transit only a very small
fraction of the stellar disk of its parent star.

The RM effect can be detected even in spectra taken with low resolution,
such that the profile changes can not be resolved. This is done by
deriving the stellar radial velocities from the spectra obtained
during the eclipse. The velocities are expected to have shifted from
their orbital motion value, because the center of gravity of the lines
are shifted by the eclipsing component. When the red-shifted part of
the stellar disk is eclipsed the velocity is shifted to the blue and
{\it vice versa}. The amplitude of the shift depends on the relative
spin-orbit inclination, but also on the rotational velocity of the
eclipsed star, the ratio of the two radii and the limb darkening of
the stellar disk (Kopal \nocite{kopal1942}1942; 1959; Hosokawa
\nocite{hosokawa1953}1953; Ohta \etal\ 2005; Winn \etal\
\nocite{winn2005}2005; Gim{\'e}nez \nocite{gimenez2006} 2006).

In fact, almost all observations of the RM effect were performed by
observing the distortion of the radial-velocity curve during the
eclipse. This includes the original work of Rossiter and
McLaughlin, and the more recent work of Worek
(\nocite{worek1996}1996), who observed AI Dra and V505 Sgr. In the next
subsection we review two additional interesting observational studies
of DE Dra and V1143 Cyg.

\subsection{Interesting cases}

\subsubsection{DE Dra}

DE Dra is an eclipsing single-lined spectroscopic binary (Furtig \&
Meinunger \nocite{furtig1976}1976), consisting of a B9V primary in a
5.3 day circular orbit (Hube \nocite{hube1976}1976). Hube \& Couch
(\nocite{hube1982}1982) obtained many spectra of the system during and
outside the eclipse, in order to derive the radial-velocity modulation
of the system. The obtained radial velocities clearly showed the RM
effect during eclipse. The distortion of the radial-velocity curve
was significantly asymmetric, suggesting a spin-orbit inclination of
the primary. The analysis of Hube \& Couch of the distorted
radial-velocity modulation indicated a rotational broadening of 180 km
s$^{-1}$. They performed an independent line profile analysis
which yielded  a rotational velocity of only 135 km s$^{-1}$. It seems as if
the inconsistency between these two values prevented Hube \& Couch from
carrying their analysis of the RM effect further and deriving a firm value
for the relative inclination of the primary axis.

The observations and analysis of Hube \& Couch (1982) were not
performed with modern techniques, so the statistical significance of
their results is questionable. However, if true, this is the only
binary that we know of that has a non-aligned rotational axis. This is
most interesting because it can tell us something about the binary
formation of DE Dra. In fact, the system has another feature which
indicates that the binary has not yet reached tidal equilibrium. The
primary rotational velocity, being either 180 or 135 km s$^{-1}$, is
spinning much faster than the synchronization velocity, which is about
30 km s$^{-1}$. Therefore, we probably have in hand a binary system
that has completed its circularization, if it was formed with
eccentric orbit, but has not yet reached synchronization and
alignment.

The exceptional features of this system make further observations
and analysis very desirable. New observations can confirm the
non-alignment of the primary, resolve the secondary spectrum,
study the secondary eclipse, and search for forced precession of the
primary, which should manifest itself in varying inclination of the
stellar axis. The latter feature would add new information that would
contribute to our understanding of this remarkable system.

\subsubsection{V1143 Cyg}
\label{V1143_RM}

Just before the conclusion of this manuscript, Albrecht \etal\
(\nocite{albrecht2007}2007) put a beautiful paper in {\it astro-ph}, which  reports on a very careful study of the RM effect in the eclipsing
binary V1143 Cyg. This binary (see Section~\ref{V1143}) consists of
two F5V stars orbiting each other with a period of 7.6 days and an
eccentricity of 0.54. As stated above, it is one of the very few
binaries of which we know from observations that one of the
components reached pseudo-synchronization for a highly eccentric
orbit. However, only the secondary reached pseudo-synchronization,
while the primary is still rotating with a period which is larger by
50\% than the pseudo-synchronization period (Andersen \etal\ 1987).
The fact that the apsidal motion of the system (Gim{\'e}nez \&
Margrave \nocite{gimenez1985}1985) {\it might} be slightly smaller
than the theoretical expectation (see discussion above in
section~\ref{DI_Her} for DI Her) motivated Albrecht \etal\
(\nocite{albrecht2007}2007) to study the relative inclination of the
two stars through the RM effect.

Analysing high S/N ratio spectra in detail, Albrecht \etal\ derived
the angle $\beta$ between the stellar spin axis and the orbital
angular momentum, both projected onto the plane of the sky. Obviously,
$\beta=0$ is consistent with perfect spin-orbit alignment.  They found
that $\beta_A=0.5\pm 4^{\circ}$ and $\beta_B=-3.9\pm 4^{\circ}$, for
the primary and the secondary, respectively, to be consistent with a perfect
alignment of both components.

These results are most interesting, showing that the primary is
aligned despite the fact that the binary is not synchronized. We would
like to suggest that this might indicate that the V1143 Cyg system was
formed with a rotation axis aligned with the orbital binary motion, at
least for the primary. This is so because we {\it assume} that the
timescale for pseudo-synchronization and alignment are similar, and
therefore the non-synchronization of the primary indicates that there
was not enough time during the lifetime of the system for the primary
to reach synchronization or alignment. If this is true, observing the
RM effect of V1143 Cyg told us something about the formation of the
binary instead of its tidal evolution.

\subsubsection{KH 15D}
\label{KH_RM}

KH 15D is a T Tau binary with a period of 48 days, which is surrounded
by a precessing circumbinary disk (e.g., Herbst \etal\
\nocite{herbst2002}2002). The primary and the secondary are being
occulted periodically by the disk (for a more detailed account of this
interesting system see Section~\ref{KH_15D}).

Winn \etal\ (2006) pointed out the possible RM effect induced by the
disk when occulting the visible star of KH 15D or its halo. However,
the optical spectra they obtained did not allow a quantitative analysis of
this effect. Therefore, there is no information available for
alignment or misalignment of the primary of KH 15D relative to its
binary orbit. However, strong evidence was found for an inclined, warped,
eccentric, precessing circumbinary disk (Chiang \& Murray-Clay; Winn
\etal\ 2004; 2006) with a precession period of the order of 1000
years.  The cause of the misalignment between the binary plane of
motion and its circumbinary disk is still a mystery, and might
indicate some basic misaligned component of the angular momentum left
from the formation era of the system.

%
\section{Tidal Interaction in Close Triple System}
\label{triple}
%

As stated in the introduction, tidal interaction in close binaries
acts until the binary reaches an equilibrium state, in which the orbit
is circularized, the stellar rotations are synchronized and the
stellar spins are aligned with the orbital motion. It is only when
this state is reached that the stellar tides do not move relative to
the surfaces of the two stars, and the stellar shapes are constant in
the rotating frame of reference.

However, when a binary is a part of a hierarchical triple system,
which is composed of the binary and a third distant companion, such a
stable configuration might not be possible.  The reason for this is that the
third star constantly injects eccentricity into the binary orbit by
exerting its own `tidal forces' on the binary system as a whole. The
origin of these tidal perturbations is the small difference between
the gravitational acceleration of the two individual stars induced by
the third star and the corresponding acceleration of the center of
mass of the binary system. This difference does not depend on the
actual radii of the two stars, but instead depends on the size of the
binary system relative to its distance from the third star.

One can classify the perturbation of the third star into three
different effects.

\begin{itemize}
\item Modulation of the binary eccentricity.

\item  Apsidal precession of the binary.

\item Nodal precession of the binary orbit around the total angular
 momentum of the triple system.
\end{itemize}
The effects of the third star can appear on three timescales: the
binary period, the third star period, and a long-term modulation
(e.g., Brown 1936; Borkovits \& Forg\'{a}cs-Dajka 2008, this
volume). In what follows we are mainly interested in the long-term
modulation and its interrelation with the long-term tidal interaction
that is occurring in the binary.

It is advisable that one take the possible effects of an unknown
distant companion into consideration when observing binary stars,
because a third distant companion could easily escape the notice of
observers in the course of studying binary systems. Unfortunately, in
many cases no effort is made to search for a faint, distant companion
after a binary is discovered and studied. Nevertheless, several lines
of evidence indicate that the percentage of binaries that have a
distant third companion is significant.

Mayor \& Mazeh (1987) were the first to search systematically for
third companions of spectroscopic binaries, using the precession of
the nodes induced by the third companion. They looked in a small
sample of spectroscopic binaries and found evidence that 25\% of the
members of their sample have third companions.
Other studies (Isobe
\etal\ \nocite{Isobe1992}1992; Tokovinin \& Smekhov
\nocite{tokovinin2002}2002; Pribulla \& Rucinski
\nocite{pribulla2006}2006; D'Angelo \etal\ \nocite{dangelo2006}2006)
also indicated a high frequency of triple systems. A recent search by
Tokovinin \etal\ (\nocite{tokovinin2006}2006) found that 96\% of
binaries with periods shorter than 3 days have a third companion! A
major step forward in the observational study of triple systems is the
construction of a new catalog by Tokovinin which
includes all known stellar triple systems
(\nocite{tokovinin1997}1997; see also the on-line version  at http://www.ctio.noao.edu/~atokovin/stars/index.php).

Note that the nodal precession of the orbital plane of a binary caused
by a third star, an effect used by Mayor \& Mazeh (1987) to search for
triple systems, might induce spin-orbit misalignment into the binary
system (Fabrycky \& Tremaine 2007)\footnote{{\tt I thank D.~Fabrycky
for pointing this point to me.}}.
Therefore, observational evidence for such misalignment might suggest
a third companion. Spin-orbit misalignment due to a third star was
suggested by Beust \etal\ (\nocite{beust1997}1997) for the eclipsing
binary TY Cor Aus.

This section will review the theory of long-term binary eccentricity modulation
and apsidal motion induced by a third star and will point out its
application for the tidal evolution of binaries that reside in triple
systems. In fact, as suggested already by Mazeh \& Shaham (1979) and
now supported by the observations and statistical analysis of
Tokovinin \etal\ (2006), a third companion might have a major role
in the formation of very short binaries through constant pumping
of the binary eccentricity.

\subsection{The binary modulations induced by a third distant companion}
\label{subsection_triple_ecc}

The zero-order approximation of the dynamics of hierarchical triple
stellar systems assumes that the close binary moves in an orbit with
its own semi-major axis $a_{1,2}$ and period $P_{1,2}$, while the
third distant companion moves together with the binary center of mass
in an orbit with $a_3$ and $P_3$. The hierarchy of the system relies
on the ratio $a_{1,2}/a_3$ being small.  Harrington
(\nocite{harrington1968}1968; \nocite{harrington1969}1969) developed a
theory of hierarchical triple stellar systems by expanding their
Hamiltonian into a series of terms in the small parameter
$a_{1,2}/a_3$. In order to obtain the long-term modulation, Harrington
further averaged the Hamiltonian over the binary and the third star
periods. The resulting Hamiltonian showed that the tidal forces of the
third star vary the binary eccentricity $e_{1,2}$, the longitude of
the periastron $g_{1,2}$ and the direction of the plane of motion with
quasi-periodic modulation. The modulation period is of the order of
$T_{mod}$:

\begin{equation}
T_{mod}=P_{1,2}\left(\frac{a_3}{a_{1,2}}\right)^3\,
\frac{M_1+M_2}{M_3} \ ,
\end{equation}
where $M_1$ and $M_2$ are the masses of the two components of the
binary and $M_3$ is the mass of the distant companion (Mazeh \&
Shaham 1979). One can then derive the equation of the binary
eccentricity:

\begin{equation}
\frac{de_{1,2}}{dt}=\frac{2\pi}{T_{mod}}\gamma \sin(2g_{1,2})e_{1,2}
\ ,
\end{equation}
where $\gamma$ is a geometrical factor of order unity (Mazeh \& Shaham
1976; 1979). This equation conjugates the binary eccentricity
variability with that of the longitude of the periastron, which is
also being varied by the third star:
\begin{equation}
\frac{dg_{1,2}}{dt}=\frac{2\pi}{T_{mod}}\gamma [\delta+\cos(2g_{1,2})]
\ ,
\end{equation}
where $\delta$ is another geometrical constant of order unity. (For a
different approach to the derivation of the apsidal motion induced by
a third star see, for example, Brown \nocite{brown1936}
\nocite{brown1937}1936; 1937). The two equations consider the three stars
as point masses and ignore general relativistic effects.

Further studies of the dynamics of triple stellar systems as three
mass points include the work of Krymolowski \& Mazeh
(\nocite{krymolowski1999}1999) and Ford \etal\
(\nocite{ford2000}\nocite{ford2004}2000; 2004) who expanded the
Hamiltonian to higher orders of $a_{1,2}/a_3$ and that of Beust
(\nocite{beust2003}2003), who applied a symplectic integration to the
stellar triple problem.

In any specific triple system, the actual binary apsidal motion
includes the precession induced by the binary tidal interaction and
the general relativistic effect (see Section 3).  Bozkurt \& De{\u
g}irmenci (\nocite{bozkurt2007}2007) have just recently published a
study of the apsidal motion of several known triple systems with
detected apsidal motion. However, for all the systems they considered,
the apsidal motion induced by the third star is several orders of
magnitude slower than the observed one.  The reason for this is that in all
systems considered by Bozkurt \& De{\u g}irmenci, the existence of the
third star was discovered by timing the eclipses minima and using the
light-travel time technique.  In principle, this is the same technique
as that by which the apsidal motion was detected. Therefore, the
orbital period of the third star, $P_{3}$, is necessarily of the same
order of magnitude as the detected apsidal motion. Consequently, the
apsidal motion induced by the third star is much longer than the
detected one. This can be seen by writing Equation (7.1) as

\begin{equation}
T_{mod} \simeq P_{3}\left(\frac{P_3}{P_{1,2}}\right)\,
\frac{M_1+M_2}{M_3} \ ,
\end{equation}
which implies that $T_{mod}$ is much longer than $P_3$. In order to find triple
systems with shorter apsidal binary motion, one need to look for triple
systems with a much smaller $P_{3}/P_{1,2}$ ratio (see Tokovinin
(1997) catalog of multiple systems). Two such systems are G:38-13
(Mazeh \etal\ \nocite{mazeh1993}1993) and HD 109648 (Jha \etal\
\nocite{jha2000}2000). In these systems the period of the
modulation induced by the third star is of the order of 100 years.

\subsection{The eccentricity modulation 
combined with tidal circularization}
\label{subsection_MS}

In many triple systems, $T_{mod}$ is of the order of the
circularization timescale, $\tau_{circ}$, or {\it shorter}.  In such
cases one must take the modulations induced by the third star into
account when considering the tidal evolution of close binaries,
particularly their eccentricity.

\begin{figure}
\includegraphics[width=13cm]{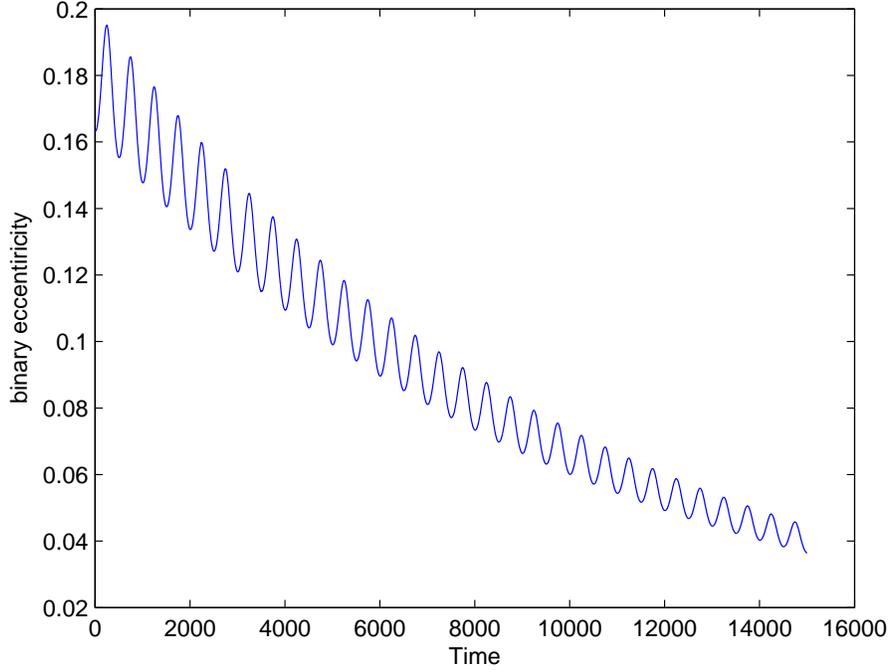}
\caption{Schematic presentation of the eccentricity evolution of a
  binary that resides in a triple system, after Mazeh \& Shaham
  (1979).}
\label{fig_ecc_evolution}
\end{figure}

Mazeh \& Shaham (1979), for example, considered the long-term effect of
the eccentricity modulation induced by the third star together with
the binary tidal circularization (Equation~\ref{eq_circ})

\begin{equation}
\frac{de_{1,2}}{dt}=-\frac{1}{\tau_{circ}}e_{1,2}
\ ,
\end{equation}
which is also linear in $e_{1,2}$, and got, to first order, a
periodic modulation combined with an exponential decay, as displayed
schematically in Figure~\ref{fig_ecc_evolution}.

Figure~\ref{fig_ecc_evolution} presents a moderate third-star
modulation of the eccentricity. However, this is not necessarily the
case for all systems, as it all depends on the relative inclination of
the two orbits. As pointed out by many studies (e.g., Krymolowski \&
Mazeh \nocite{krymolowski1999}1999; Ford \etal\
\nocite{ford2000}\nocite{ford2004}2000; 2004) the third star 'tidal
forces' can inject a high eccentricity into the binary motion, if
the inclination angle of the plane of motion of the third star
relative to the binary plane of motion is higher than approximately
$40^{\circ}$. This high-amplitude modulation was first noted by Kozai
(\nocite{kozai1962}1962) for the motion of asteroids and was first applied
to stellar systems by Mazeh \& Shaham (\nocite{mazeh1976}1976),
who at that time were unaware of Kozai's work.

As the third star modulation and the tidal circularization are both
linear in $e$, one {\it could} conclude that at the end of the
circularization process the modulation decays to zero. However, Mazeh
and Shaham (1979) have demonstrated that the linearity of the third star
modulation is true only to the first order of $a_{1,2}/a_3$. Upon
integrating Newton's equations directly, they discovered that the
eccentricity modulation is present even if the binary eccentricity
starts with zero. The Mazeh \& Shaham effect was further theoretically
studied and confirmed by Georgakarakos \nocite{georgakarakos2002}
\nocite{georgakarakos2004}(2002; 2004). We therefore expect some
short-period binaries that should have been circularized long ago to
display small eccentricities. 

In fact, observational evidence for such an effect has already been
pointed out by Mazeh (\nocite{mazeh1990}1990) in three short-period
binaries, with significant eccentricities {\it and} strong indications
for a third distant stellar companion. One of the examples found by
Mazeh is the Hyades star BD $+23^{\circ}635$, which has an orbit of
2.4 days and an eccentricity of $0.057\pm0.005$ (Griffin \& Gunn
\nocite{griffin1981}1981). Giuricin \etal\ (1984) also suggested that
the small but significant eccentricities they found in systems with
fractional primary radius larger than 0.3 (see
Section~\ref{subsection_early_type} for details) must be due to
unknown third distant companions. In all these cases, the tidal
processes should have circularized the orbits long ago, and the
interaction with the third star seems to be the only plausible
conjecture that accounts for the observed eccentricity.

The discovery of Mazeh \& Shaham (1979) that the binary eccentricity
induced by the third star continues its modulation even after the
complete tidal circularization may have implications for the evolution
of close binaries. The eccentricity modulation implies that binaries
with close enough companions will never get into a stable circular
orbit, which means that the binary tidal dissipation will continue to
work forever.  The injected binary eccentricity invokes frictional
forces inside the two stars that dissipate rotational energy and
transfer angular momentum between the stellar rotation and binary
orbital motion. This causes the semi-major axis of the binary to
decrease at a rate of
\begin{equation}
\frac{1}{a_{1,2}}\frac{da_{1,2}}{dt}=-\frac{1}{T_{circ}}Ae^2_{1,2} \ ,
\label{eq_da_dt}
\end{equation}
where $A$ is a geometrical factor of order unity.  Mazeh and Shaham
(1979) suggested, therefore, that in certain configurations the third
distant companion can provoke the spiral-in of the close binary at a
rate that can be derived by averaging Equation~\ref{eq_da_dt}.  The 
third star motion can serve as the sink for the angular momentum of
the shrinking binary.

Obviously, the separation of the perturbations of the third star and
the tidal friction into two equations is artificial, and was done only
as a first approximation.  Further studies, including Kiseleva \etal\
(\nocite{kiseleva1998}1998), Eggleton \& Kiseleva-Eggleton
(\nocite{eggleton2001}2001; \nocite{eggleton_kis2006}2006), Eggleton
(\nocite{eggleton2006}2006), Borkovits \etal\
(\nocite{borkovits2004}2004; \nocite{borkovits2005}2005;
\nocite{borkovits2007}2007) and Fabrycky \& Tremaine
(\nocite{fabrycky_tremaine2007}2007) combined the two effects into one
set of coupled equations. Kiseleva-Eggleton, Eggleton and Fabrycky \&
Tremaine further studied the Mazeh \& Shaham effect for cases where
the eccentricity modulation is large, which occurs when the relative
angle between the two orbital planes of motion is larger than $\sim
40^{\circ}$. They also concluded that some of the presently close
binaries could have been formed by the combination of tidal binary
friction and the third star induced modulation. The Tokovinin \etal\
(2006) result, that 96\% of binaries with periods shorter than 3 days
have a third companion, might be a strong evidence for the Mazeh \&
Shaham conjecture.

%
\section{Extrasolar planets and tidal Interaction}
%

Since the discovery of 51 Peg b (Mayor \& Queloz
\nocite{mayor1995}1995), more than 200 planets have been discovered
(see, for example, the Extrasolar Planets Encyclopaedia\footnote{{\tt
http://vo.obspm.fr/exoplanetes/encyclo/catalog-RV.php}}).  As one can
consider a star with a planet as a binary with an extremely small mass
ratio, we can apply the theory of tidal interaction to these systems,
and therefore expect to find some signs of its action. In particular,
we expect the planets with short enough orbital periods to attain
circularization, alignment, and planetary synchronization. This
section will review in brief some observational lines of evidence that
might indicate that tidal interaction between extrasolar planets and
their parent stars took place during the lifetime of those systems.

\subsection{Radial drift by tidal interaction}

One possible outcome of the tidal interaction between a planet and its
parent star is a radial drift of the planet, resulting from the
transformation of angular momentum between the stellar rotation and
the planetary orbital motion. The angular momentum of the planetary
orbital motion is proportional to $m_p\sqrt{a_p}$, where $a_p$ is the
planetary orbital radius and $m_p$ is the planetary mass, and is small
relative to the stellar angular momentum for planets with orbital
periods of a few days.  Therefore, transfer of a small amount of
angular momentum from or into the planetary orbital motion can
substantially change the planetary orbital radius, an effect which is
unique to planetary motion. This subsection discusses the two possible
directions of the radial drift and their astrophysical implications.

\subsubsection{Stopping the migration}
\label{subsubsection_stopping}

One possible effect of the radial drift was suggested already by Lin
\etal\ (\nocite{lin1996}1996) for 51 Peg b, whose orbital radius is
about 0.05 AU. The idea is based on the paradigm that Jupiter-sized
planets can not be formed close to their parent stars, at a distance
of $\sim0.05$ AU, but instead must be formed behind the 'ice line' of
the protoplanetary disk, at a distance of $\sim5$ AU (e.g.,
Bodenheimer \& Pollack \nocite{bodenheimer1986}1986; but see, however,
Boss \nocite{boss1997}1997 for a heretical approach). At that distance
the icy particles initially form a rocky core of several earth masses,
which subsequently accretes gas to form a giant planet.  In order to
save the core-accretion model, Lin \etal\ suggested that 51 Peg b
migrated from the location of its formation into its present radius
through interaction with the material in the disk, an idea that had
already been discussed a decade before the discovery of 51 Peg b
(e.g., Goldreich \& Tremaine \nocite{goldreich1980}1980; Papaloizou \&
Lin \nocite{papaloizou1984}1984; Lin \& Papaloizou
\nocite{lin1986}1986) in the context of the solar system. Such a model
for the short-period extrasolar planets had to explain why the
migration stopped at $\sim0.05$ AU and did not push the planet into
the surface of the star. The idea of Lin \etal\ was that this was done
by the tidal interaction between the planet and its parent star.

This 'stopping mechanism' can work only if the star rotates on its spin
axis faster than the orbital frequency of its planet. In such
a case, the tidal interaction between the planet and the star tends to
slow down the stellar rotation, taking angular momentum from the star
into the planetary orbital motion. The additional orbital angular
momentum forces the planetary orbital radius to {\it increase}. This
result is counterintuitive, because the tidal interaction eventually
{\it slows} both the planetary orbital motion and the stellar
rotation. Nevertheless, it relies on the conservation of angular
momentum and Kepler's laws, and must therefore be correct.

According to this scenario, the planet can migrate inwards as long as the
interaction with the disk is stronger than the stopping force of the
tidal interaction with the star. The planet stops its migration when
the two forces cancel each other out.  The timescale of the
radial drift caused by the interaction with the spinning star was
estimated by Lin \etal\ (1996) to be:
\begin{equation}
\tau_{radial}=\frac{a_p}{\mid\dot{a}_p\mid}=\frac{9}{2n_p} q^{-1}
\left(\frac{a_p}{R_*}\right)^5 Q_* 
\propto \frac{M_*^{1/2}Q_*}{R_*^5} \, \frac{a_p^{13/2}}{m_p}
\label{radial_timescale}
\end{equation}
where $n_p$ is the Keplerian angular velocity of the planet,
$q=m_p/M_*$ is the mass ratio between the planetary mass, $m_p$, and
the stellar mass, $M_*$, $R_*$ is the stellar radius, $a_p$ is the
planetary orbital radius and $Q_*$ is the stellar tidal dissipation
factor (See Sasselov \etal\ 2003 for a different expression of
$\tau_{radial}$). Note that Trilling \etal\
\nocite{trilling1998}(1998) pointed out that the numerical coefficient
in this equation is only true for bodies of uniform density, and
should be corrected for the actual stellar mass profiles.

The problem with this equation is the unknown factor $Q_*$, which
describes how efficiently rotational energy is dissipated by friction
within the star. Lin \etal\ (1996) adopted a value of $1.5\times 10^5$
``for a main-sequence star, based on the observations that the orbits
of short-period pre-main-sequence binary stars and the main-sequence
binary stars in the Pleiades cluster are circularized for periods up
to 5 and 7 days, respectively''. We note, however, that the error on
this derivation of the value of $Q_*$ from the transition period of
the Pleiades could be substantial for two reasons. First, the binaries
in the Pleiades cluster might have been circularized during their
pre-main-sequence phase. In such a case, we can not use the transition
period of the Pleiades at all. Second, we can not estimate the exact
transition period, nor do we know the exact mass of the secondaries in
the binaries of those clusters, as was discussed in
Section~\ref{subsub_transition} in detail. The factor $Q_*$ is also a
strong function of the star itself, depending on the depth of the
convective envelope and the stellar radius at the time of the inward
migration (e.g., Sasselov 2003).

As we can see from Equation~\ref{radial_timescale}, the stopping
mechanism is a strong function of the planetary orbital radius, and
the radial-drift timescale drops dramatically for small distances. The
equilibrium between the interaction with the disk and the tidal
interaction with the spinning star depends on the disk mass and
density profile on the one hand, and the stellar structure and the
intensity of the stellar interaction on the other hand. According to
this scenario, when the protoplanetary disk disappears, the planet will
drift outward until its radial-drift timescale gets too long.

\begin{figure}
\includegraphics[width=13cm]{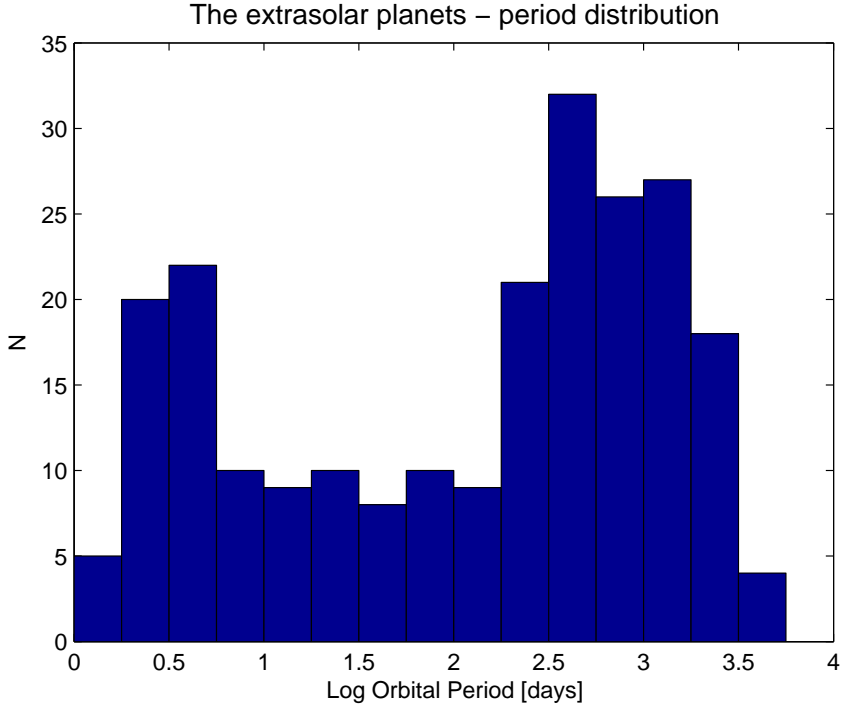}
\caption{The period distribution of the extrasolar planets.}
\label{extrasolar_period}
\end{figure}

Some support for the stopping mechanism scenario can be found in the
period distribution of the known planets, which is plotted in
Figure~\ref{extrasolar_period} (Data were taken from the Extrasolar
Planets Encyclopaedia\footnote{{\tt
http://vo.obspm.fr/exoplanetes/encyclo/catalog-RV.php}}).  One
possible interpretation of the figure is that between 2 and 200 days
the distribution is flat in log period, with an excess of about 20
planets around the period of 3 days. According to the migration
scenario and its stopping extension, at least half of the planets with
periods of around 3 days were stopped by the tidal interaction during
their inward migration. The excess of planets with periods around 3
days is still there even if we take out the transiting planets, for
which there is a strong selection effect at short periods. 

This interpretation of the observed distribution implies that many of
the parent stars of the planets with orbital periods of 3 days have,
or at least had in the past, rotational periods shorter than 3 days.
We can try to corroborate this conjecture by deriving the present
stellar rotational periods of these systems.  One approach to deriving
the rotation period is based on the rotational broadening of the
stellar profiles, relying on the stellar radius and
inclination. However, even if we know the inclination, this approach
is quite inaccurate, as discussed in Section~\ref{section_synchro}.

Another approach is to follow the photometric variation of the parent
stars that can reveal their rotational periods. Consider, for example,
the M-type star Gliese 436, whose planet was discovered by Butler \etal\
(\nocite{butler2004}2004), and its transiting nature was revealed only
recently by Gillon \etal\ (\nocite{gillon2007}2007). Demory \etal\
(\nocite{demory2007}2007) found some evidence for a photometric
modulation with a period of about 45 days, which could be the
rotational period of the parent star. If this is true, then the
rotational period of the M star is much longer than the 2.6 day
orbital period of the planet.  Bonfils \etal\
(\nocite{bonfils2007}2007) found a similar situation for Gliese 674,
for which the orbital period is 4.7 days and the rotation period of
the M star is about 35 days. However, the assumed long rotational
periods of these two stars could be the result of spinning down of
the stellar rotation, a commonly observed phenomenon, in M-type stars in
particular (e.g., Cardini \& Cassatella \nocite{cardini2007}2007).

Other mechanisms can be invoked to explain the peak of the period
distribution of the extrasolar planets at around 3 days. For example,
one can assume that the protoplanetary disks have been evaporated when
the planets arrived into their present radii, which correspond to
periods around 3 days.  After all, we know that the disks disappear
after a period of the order of 10 Myrs, either by stellar wind and
photoevaporation or by viscous processes in the disk itself (e.g.,
Haisch \etal\ \nocite{haisch2001}2001).  Another version of this
stopping mechanism was suggested by Kuchner \& Lecar
(\nocite{kuchner2002}2002), who assumed a central 'hole' in the
protoplanetary disks, with an edge corresponding to an orbital period
of 6 days. The central hole is formed by turbulent accretion due to
magnetorotational instability that acts on the material in the central
part of the disk, where small grains reached sublimation
temperature. The planetary migration halts when the exterior 2:1
Lindblad resonance reaches the external radius of the hole.  Another
stopping mechanism (e.g., Trilling \etal\ 1998) is based on the
assumption that the short-period planets got so close to their parent
stars that their Roche Lobe became smaller than their planetary
radii. This caused mass transfer from the planets onto the stars,
which resulted in an outward motion of the planet in order to conserve
angular momentum.

A different idea was raised recently by Burkert \& Ida
(\nocite{burkert2007}2007), who suggested that the period distribution
peak was in fact a pronounced dip at the range of 10--100
days. Furthermore, their analysis indicated that the dip appeared only
in planets around F-type stars, with masses larger than $1.2
M_{\odot}$, while the distribution of planets around G-type stars showed
log-flat distribution. To account for the different period
distributions, Burkert \& Ida assumed that the sizes of the
protoplanetary disks around F and G stars were different, resulting in
 different patterns of migration.  Eventually, this
caused a gap in the period distribution of planets around F
stars, but not around G stars. It seemed that their simulations could
produce the period distribution of the planets and did not necessitate
an extra stopping mechanism.

We note that the distinction between the period distribution of
planets around F and G stars is an interesting feature of the planet
population, and might be accounted for by different tidal interactions
with the parent stars, instead by the different masses of the
disks. This is so because tidal interaction in F stars is
substantially weaker than in G stars, due to the fact that F stars do
not have convective envelopes. Thus, F stars can stop the inward
migration only when the planet is substantially closer to the parent
star than the stopping radius of G stars. Consequently, if tidal
interaction with their parent stars plays a role in the final orbits
of planets, the difference in the period distribution of planets
around G and F stars {\it could} have been caused by internal stellar
structure differences.

The present discussion clearly demonstrates that the extra planets at
an orbital period of around 3 days are still not acceptable evidence
for the action of tidal interaction between short-period planets and
their parent stars, and that further observations and theoretical work
are still necessary.

\subsubsection{Inward drift}

Tidal interaction can also push the planet inwards, if the planet
orbital frequency is faster than the stellar spin. In such a case, the
planet is dragged back as a result of tidal interaction with the star, an
interaction that takes angular momentum from the planetary orbital
motion, which results in a decrease in the orbital radius of the
star. The timescale for the inward drift is given by
Eq.~\ref{radial_timescale} that describes the outward drift, but now
$\dot{a}_p$ is negative. However, unlike the outward drift, the inward
drift is a runaway process, in which $a_p$ gets smaller and therefore
the drift timescale, $\tau_{radial}$, gets shorter, which accelerates the
drift further (e.g., Rasio \& Ford \nocite{rasio1996b}1996).

It is therefore convenient to define the corotation radius, in which
the planetary orbital period equals the stellar rotational
period. Outside this radius the planet is pushed outward, while within
this radius the planet is pushed inward. If interaction with the disk
succeeds in pushing the planet inside the corotation radius, then
tidal interaction with the star joins forces with the disk interaction
to push the planet inward. The planet can survive inside its
corotation radius only if the disk disappears {\it and} the drift
timescale is long enough. Rasio \& Ford concluded that for 51 Peg the
timescale for the inward drift is shorter than the stellar lifetime
only if the orbital period is shorter than about 10 hours. This means
that we would not be able to find planets around stars similar to 51
Peg with orbital periods shorter than 10 hours, which indeed is the
case.

We note that the stellar radii and rotational periods do change during
the stellar lifetime, and both the corotation radius and the
radial-drift timescale necessarily follow these variations. As the
stellar rotational period gets longer and the corotation radius
increases, the planet can find itself inside the corotation radius,
being pushed inwards by the tidal interaction with its parent
star. Such a planet can survive only if its orbital radius is large
enough so the drift timescale is longer than the stellar
lifetime. 

As of November 2007, eleven planets were detected with periods shorter
than 2 days, the shortest of which is OGLE-TR-56b, with a period of
1.2 days (Konacki \etal\ \nocite{konacki2003}2003). The discussion
above highlights the need to carefully consider the stability of the
orbits of these planets against inward drift. Sasselov (2003), for
example, tried to exclude some versions of the tidal theories that
predict too fast radial drift for OGLE-TR-56b. He suggested that for
some theories it would be possible to detect in the near future
changes in the orbit of OGLE-TR-56b by precise timing of its transit.
Such a measurement would be a landmark of the observational evidence
for tidal interaction in main-sequence stars, enabling us to follow
the results of the interaction in real time, and not only collect
indirect proofs for its action in the past.

\subsection{Circularization of extrasolar-planet orbits}

\begin{figure}
\includegraphics[width=13cm]{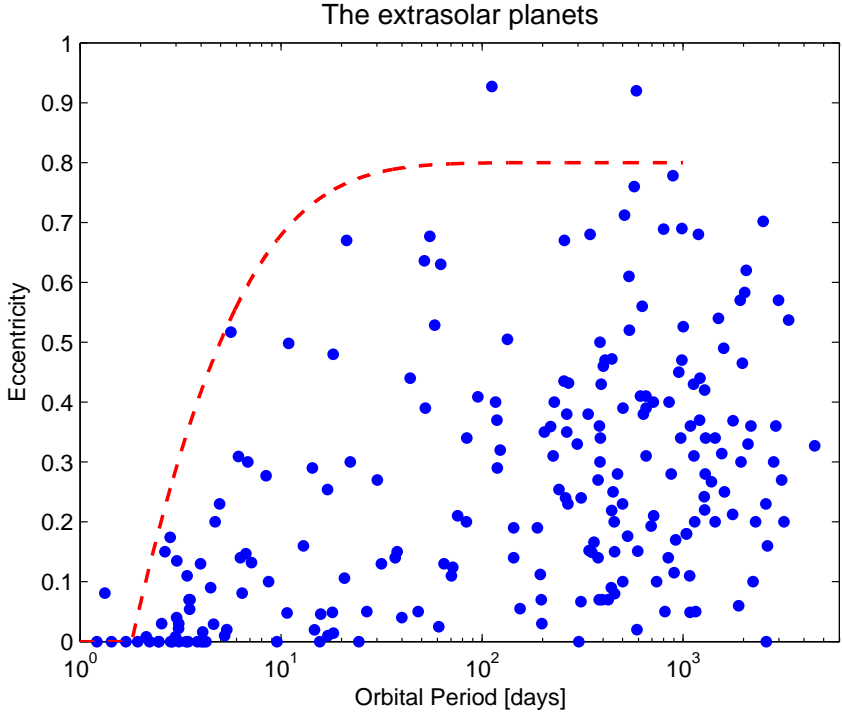}
\caption{The eccentricity of the extrasolar planets as a function of
  the orbital period. The dashed line is the upper envelope, with an
  equation $f=E-A \exp(-(p B)^c)$, where $E=0.8$, $A=8$, $B=6$ and
  $C=0.35$.}
\label{extrasolar_ecc}
\end{figure}

In order to study the tidal circularization of extrasolar planets, we
plotted in Figure~\ref{extrasolar_ecc} the eccentricities versus
periods of all known planets, like we did in Figure~\ref{e_log_p} for
all the known spectroscopic binaries. The main two features of
Figure~\ref{e_log_p} can also be found in Figure~\ref{extrasolar_ecc}
(see Halbwachs \etal\ \nocite{halbwachs2005}2005 for a detailed
comparison). All planets with periods shorter than a 'cutoff' period
are circular, and all the extrasolar planets with longer periods have
eccentricities below an 'upper envelope', which starts at eccentricity
zero and climbs up asymptotically to a certain high value. The values
of the planetary upper envelope parameters are different from those of
the spectroscopic binaries. The upper envelope starts to rise at a
period of 2.5 days and rises to an eccentricity of 0.8 when moving
towards 25 days.

Figure~\ref{extrasolar_ecc} displays only three exceptions to these two
features. Two of them are the very highly eccentric planets HD
80606b (Naef \etal\ \nocite{naef2001}2001) and HD 20782b (Jones \etal\
\nocite{jones2006}2006), both with an eccentricity of 0.92, and the
third one is HD 41004Bb --- a planet with one of the shortest periods,
1.3 days, with a small but significant eccentricity of $0.081\pm0.012$
(Zucker \etal\ \nocite{zucker2003}2003; \nocite{zucker2004}2004). Note
that the three eccentricities would not be considered exceptions
for spectroscopic binaries, because the upper envelope for
binaries reaches a value of 0.98, and the 'cutoff' period is as short
as 0.35 days.

HD 41004 is an interesting system which is composed of a double visual
star, HD 41004 A and B, separated by about 20 AU. Each of the two
stars has a low-mass companion. Component A has a planet with a
period of about 1000 days, with a minimum mass of 2.5 Jupiter mass
(=$M_J$), whereas component B has a companion with a short period and
a small but significant eccentricity, with a minimum mass of 18 $M_J$
(Zucker \etal\ 2004). We note that the general consensus puts the
upper-mass limit for planets around 10--15 $M_J$, and therefore it is
not clear if HD 41004Bb is a planet, or rather a low-mass companion
which should be referred to as a brown-dwarf secondary.  If HD 41004Bb
is a brown-dwarf companion, it should be moved from
Figure~\ref{extrasolar_ecc} to Figure~\ref{e_log_p}, where its
location is below the upper envelope. Another possibility is that the
visual companion, HD 41004A, pumped eccentricity into the orbit of Bb,
as is discussed regarding triple systems in
Section~\ref{subsection_triple_ecc}. A similar idea was already
proposed for another high eccentricity planet, 16 Cyg B (Mazeh \etal\
\nocite{mazeh1997}1997; Holman \etal\ \nocite{holman1997}1997). For a
review of this idea see Mazeh (\nocite{mazeh2007}2007).

The two prevailing features of Figure~\ref{extrasolar_ecc} might be
attributed to tidal circularization acting on the extrasolar planets,
exactly as the general consensus interprets Figure 1. However,
concealed behind the model for the eccentricity of the spectroscopic
binaries is the assumption that all binaries, without any period
difference, have been formed with the same eccentricity distribution,
including the highly eccentric binaries. Therefore, the fact that
short-period binaries are all circular must be the result of some
circularization processes occurring {\it after} the binary
formation. On the other hand, the prevailing assumption for planets
was, until just a few years ago, that all planets should have circular
or at least almost circular orbits, like all the planets in our solar
system. If planets are formed in an accretion disk, then the 'natural'
outcome is planets with circular orbits, as the interaction of the
planetesimal with the gas and dust in the disk should dissipate any
initial eccentricity. This is one of the reasons why HD114762b,
detected already in 1989 (Latham \etal\ 1989) \nocite{latham1989} with
an eccentricity of 0.4, was not considered as a planet candidate at
that time. Therefore, the discovery of planets with large
eccentricities was a surprise to the astronomical community. Thus,
theoretical effort was {\it not} devoted to explaining the circular
short-period planets, but instead to building a reasonable scenario
for the unexpected eccentricities of the long-period planets.

Some studies associated the observed eccentricities of extrasolar
planets with another surprising feature of the planets --- their small
semi-major orbital axis. As discussed in the previous subsection, the
assumed paradigm to explain the short distances of the short-period
extrasolar planets is an inward migration driven by interaction with the
disk. The effect of this interaction on
the planetary orbital eccentricity (e.g., Goldreich \& Sari
\nocite{sari2003}2003) is still a matter of controversy (see
discussion and short review by Moorhead \& Adams
\nocite{moorhead2007}2007). Other studies tried to explain the
existence of short-period planets and the planetary eccentricities as
a consequence of interaction with other, still undetected, planets
(e.g., Weidenschilling \& Marzari \nocite{marzari1996}1996; Rasio \&
Ford \nocite{rasio1996}1996; Zhou \& Lin \nocite{zhou2007}2007; Juric
\& Tremaine \nocite{juric2007}2007).  It is not clear if these
scenarios could explain the distribution of eccentricities in
long-period planets and the circular orbits of short-period planets at
the same time. If this is not possible, and the assumed mechanism that
pumped eccentricities into the long-period orbits also necessarily
built high eccentricity in some of the short-period planets, then we
might need to invoke tidal circularization to account for the circular
orbits of the short-period planets.

The circularization of a planetary orbit can come from
dissipative processes that take place either in the star or in the planet.
The circularization timescale due to processes in the {\it planet} is
usually written (e.g., Mardling 2007) as

\begin{equation}
\tau_{circ,p}=\frac{2}{21n_p} \left(\frac{Q_p}{k_p}\right)
 q  \left(\frac{a_p}{r_p}\right)^5
\ ,
\end{equation}
where $Q_p$ is the planetary dissipation parameter, $k_p$ is the
planetary tidal Love number (Goldreich \& Soter \nocite{gold1966}1966)
and $r_p$ is the planetary radius (See Mardling
\nocite{mardling2007}2007 for an essential discussion of the slightly
different definitions of the Q parameters by the different studies).
For the corresponding timescale due to processes in the star,
$\tau_{circ,*}$, one should replace $Q_p$, $k_p$, $r_p$ by the
corresponding values of the star, $Q_*$, $k_*$, $R_*$, and switch
between $m_p$ and $M_*$ (e.g., Carone \& P{\"a}tzold
\nocite{carone2007}2007).  Unfortunately, as was already noted in the
previous subsection, the theoretical estimate of
the ratios $Q_p/k_p$ or $Q_*/k_*$ varies over one or two orders of
magnitude (e.g., Lin \etal\ 1996; Trilling \etal\
\nocite{trilling1998}1998; Trilling \nocite{trilling2000}2000).  For
example, Trilling (2000) used $Q_* \geq 10^6$ for main-sequence
stars and $1.5\times 10^4$ -- $1.5\times 10^4$ for PMS
stars.  Therefore, the theory is in a 'safe' situation, in which it can
not be confronted with the observations.

\subsubsection{Small eccentricity induced by a second planet}
\label{subsection_small_e}

In Section~\ref{subsection_MS} we discussed the Mazeh \& Shaham (1979)
effect, in which a third distant companion can induce eccentricity
in the binary orbit, even for a short-period orbit for which the
tidal circularization had reduced the binary eccentricity. It was
emphasized that the Mazeh \& Shaham effect can work even if the binary
eccentricity starts at exactly zero value. This effect can have
an impact  on the evolution of close binaries, as the injected binary
eccentricity invokes frictional forces inside the two stars that
dissipate rotational energy and transfer angular momentum between the
stellar rotation and the binary orbital motion.

Similar ideas have been proposed with regard to extrasolar planets. It
was suggested that another, as-of-yet undetected planet pumps small
eccentricity into the orbit of some known short-period planets.  One
such example is HD 209458b (Mazeh \etal\ \nocite{mazeh2000}2000), the
first transiting planet (Charbonneau \etal\ \nocite{charbonneau2000}2000;
Henry \nocite{henry2000}2000) that was found to have a somewhat
surprisingly large radius (e.g., Burrows \etal\
\nocite{burrows2000}2000). The conjecture was that a small
eccentricity of HD 209458b orbit (e.g., Laughlin \etal\
\nocite{laughlin2005}2005) causes tidal dissipation inside the planet,
which serves as another source of energy that inflates the planet
(e.g., Bodenheimer \etal\ \nocite{bodenheimer2001}2001; Bedenheimer
\etal\ \nocite{bodenheimer2003}2003; Gu \etal\ \nocite{gu2004}2004;
Mardling 2007). Two other examples are WASP-1b (Cameron \etal\
\nocite{cameron2007}2007) and HAT-P-1b (Bakos \etal\
\nocite{bakos2007b}2007b).

The idea of a planetary eccentricity pumped by another as-of-yet
undetected planet can be verified or refuted by further
observations. Additional accurate radial-velocity observations can
constrain the orbital eccentricity of the known short-period planets
as well as detect unknown additional planets. Another avenue for
accurate derivation of orbital eccentricity of the transiting
planets is to detect the secondary eclipse, an approach recently
carried out in the IR by the {\it Spitzer} satellite (e.g., Deming
\etal\ \nocite{deming2005}2005). When the orbital parameters of those
planets are determined, we will be able to assess the validity of the
idea of the pumping eccentricity, particularly for transiting planets
for which we can determine the radius and eccentricity.

\subsection{Synchronization of planetary rotation}
 
Rotational synchronization of stars with short-period planets was not
considered by the astrophysical community, as it was assumed that the
masses of the planets are too small to synchronize the stellar rotation
with the planetary orbital frequency. See, for example, the discussion
of the stopping mechanism (Section~\ref{subsubsection_stopping}),
where we reviewed two cases with stellar rotational periods much longer
than the orbital periods. At the same time, it was taken for granted
that tidal dissipation inside the short-period planets has synchronized
their own rotation with their orbital frequency (e.g., Rasio \etal\ 1996;
Ogilvie \& Lin \nocite{ogilvie2004}2004).

Lubow \etal\ (\nocite{lubow1997}1997) claimed that even a planet with
a period of 200 days should be synchronized in 10 Gyrs (see also
Ivanov \& Papaloizou 2007 and Ogilvie \& Lin
\nocite{ogilvie2007}2007). As is commonly known, we find many planets
with eccentric orbits at that period (see, for example,
Figure~\ref{extrasolar_ecc}). Therefore, we expect to have planets
with orbital periods shorter than 200 days with
pseudo-synchronized rotational periods (see detailed derivation of the
pseudo-synchronization period by Ivanov \& Papaloizou
\nocite{ivanov2007}2007).

How can we observe planetary synchronization or
pseudo-synchronization?  Planets that are rotationally locked in a
circular orbit have one hemisphere exposed to intense stellar
insolation all the time, causing a significant difference in
temperature between their day and night sides. The infrared emission
of the day side was already detected by {\it Spitzer} through the
secondary eclipse of HD 209458b (Deming \etal\
\nocite{deming2005}2005; 2007\nocite{deming2007}), TrES-1 (Charbonneau
\etal\ \nocite{charbonneau2005}2005), 189733b (Deming \etal\
\nocite{deming2006}2006), and Gliese 436b (Deming \etal\
\nocite{deming_436_2007}2007; Demory \etal\ 2007), and even by the
phase modulation of the non-transiting planet around $\upsilon$ And
(Harrington \etal\ \nocite{harrington2006}2006).  The exact
periodicity and the phase of the IR modulation can confirm the
synchronization assumption (e.g., Seager \etal\
\nocite{seager2005}2005). Furthermore, IR modulation can reveal
a signature of pseudo-synchronized planetary rotation.

As-of-yet, {\it Spitzer} detected IR modulations only of planets with
circular or almost circular orbits with short periods, of the order
of 3 days.  Very recently the transit of a rather long-period
eccentric planet, HD 17156b, has been discovered (Barbieri \etal\
\nocite{barbieri2007}2007). It would be quite interesting to follow
the IR modulation of this planet and find out if the planetary
rotation is pseudo-synchronized with the orbital period of 21.2 days
and with an eccentricity of 0.67.

\subsection{Alignment of planetary motion with stellar rotation}

The {\it a priori} approach to extrasolar planetary alignment was
similar to the early attitude to planetary eccentricity. Based on the
solar system features, it was expected that the planetary orbital
plane of motion would be aligned with the stellar rotational axis. The
theory predicted that the dissipative processes in the protoplanetary
disk tend to damp the initial inclinations of the planetesimals (e.g.,
Ward \& Hahn \nocite{ward1994}1994; see also a later study by
Cresswell \etal\ \nocite{cresswell2007}2007). However, in the spirit
of the new approach to the formation of planetary eccentricities by
the interaction with the disk, Thommes \& Lissauer
(\nocite{thommes2003}2003), for example, suggested that when more than
one planet is present in the system, the interaction with the material
of the disk and the resonant interaction between the planets can turn
the system noncoplanar.  Interestingly, spin-orbit misalignment might
cause tidal heating of the planet and contribute to its inflated
radius, as was suggested by Winn \& Holman (\nocite{winn_holman2005}2005)
when trying to account for the large radius of HD 209458b (see
Section~\ref{subsection_small_e}). However, this idea was challenged
by Levrard \etal\ (\nocite{levrard2007}2007) and Fabrycky \etal\
(\nocite{fabrycky2007}2007).

To observe the relative inclination of the planetary spin axis, we can
use the RM effect for transiting planets (see Gaudi \& Winn
\nocite{gaudi2007}2007 for a thorough discussion of the effect and
Section~\ref{rm_effect} for a discussion of the same effect observed
in eclipsing binaries). The amplitude of the radial-velocity
modulation of the RM effect is
\begin{equation}
\Delta_{RM} \sim V_*\sin i_* \left(\frac{r_p}{R_*}\right)^2 \ ,
\end{equation}
where $ V_*\sin i_*$ is the rotational broadening of the stellar
spectral lines. The RM amplitude of transiting planets is necessarily
small, as $r_p/R_*\sim0.1$. It can be as small as $\sim 10$ m
s$^{-1}$, if $V_*\sin i_*$ is 1 km s$^{-1}$. Nevertheless, the effect
can have an amplitude as large as $\sim 50$ m s$^{-1}$, if $V_*\sin
i_*$ is 5 km s$^{-1}$. Such amplitude can be easily observed with
the typical precision achieved for the study of extrasolar planets.

Recent observations (e.g., Queloz \etal\ \nocite{queloz2000}2000; Winn
\etal\ \nocite{winn2006}2006b; Narita \nocite{narita2007}2007) that
aimed specifically to follow the RM effect in a few transiting planets
could not find any evidence for spin-orbit misalignment. This is true
even for 147506b (HAT-P-2b; Winn \etal\ \nocite{winn2007}2007;
Loeillet \etal\ \nocite{loeillet2007}2007) for which a substantial
eccentricity has been detected (Bakos \etal\
\nocite{bakos2007}2007a). Unfortunately, as discussed above, the
theoretical interpretation of these findings is not clear. It could be
attributed to the formation processes --- planets are formed in
aligned orbits, or it could be the outcome of some tidal dissipation
between the spinning star and the planetary motion. We are still
awaiting the discoveries of misaligned planets, which will help to
clarify the broad scope of planetary alignment.

%
\section{In the era of large-scaled photometric surveys}
%

The advent of sensitive, large CCDs in the service of the astronomical
community in the last decade or so has made large-scaled photometric
surveys possible, yielding hundreds of accurate photometric
measurements of many thousands of stars.  Although the drive for these
surveys was not always the study of binary stars, and eclipsing
binaries in particular (e.g., MACHO, see Alcock \etal\ 1977), hundreds
of eclipsing binaries have been discovered as a result of these
surveys (e.g., Devor 2005\nocite{devor2006}).

In the past, even when a star was discovered to be an eclipsing
binary, a special, additional observational effort was needed in order
to acquire its lightcurve in order to derive its orbital
parameters. In contrast, for eclipsing binaries discovered by large
surveys we get, without any additional observational effort,
lightcurves of unprecedented precision that are spread over a long
period of time. Therefore, the present large-scaled photometric
surveys are revolutionizing the study of eclipsing binaries, yielding
new tools for the study of tidal interaction in particular. This
section reviews two studies that have used two large surveys, MACHO
and OGLE, which performed an intensive monitoring of the LMC for a few
years. Both MACHO (Alcock \etal\ \nocite{alcock1997}1997; Derekas
\etal\ \nocite{derekas2007}2007)) and OGLE (Udalski \etal\
\nocite{udalski1998}1998; Wyrzykowski \etal\ \nocite{wyr2004}2004)
data sets were used to identify the largest sample of eclipsing
binaries ever studied, with direct implications for the study of their
tidal interaction. The goal of this section is to demonstrate the
far-reaching potential of these surveys.

\subsection{Apsidal motion of eccentric binaries in the LMC}

Michalska \& Pigulski \nocite{michalska2005} (2005) studied the
discovered binaries in the LMC by the OGLE survey, and searched for
binaries that are suitable for distance
determination. Serendipitously, they found 14 binaries with
significant apsidal precession. The LMC data spanned 2000 days,
and therefore they could detect slow variation of the longitude of the
periastron and derive long apsidal periods, of the order of thousands
of years. In a follow-up paper, Michalska (\nocite{michalska2007}2007)
found evidence for apsidal motion in the data of an additional 11
binaries.

\begin{figure}
\includegraphics[width=13cm]{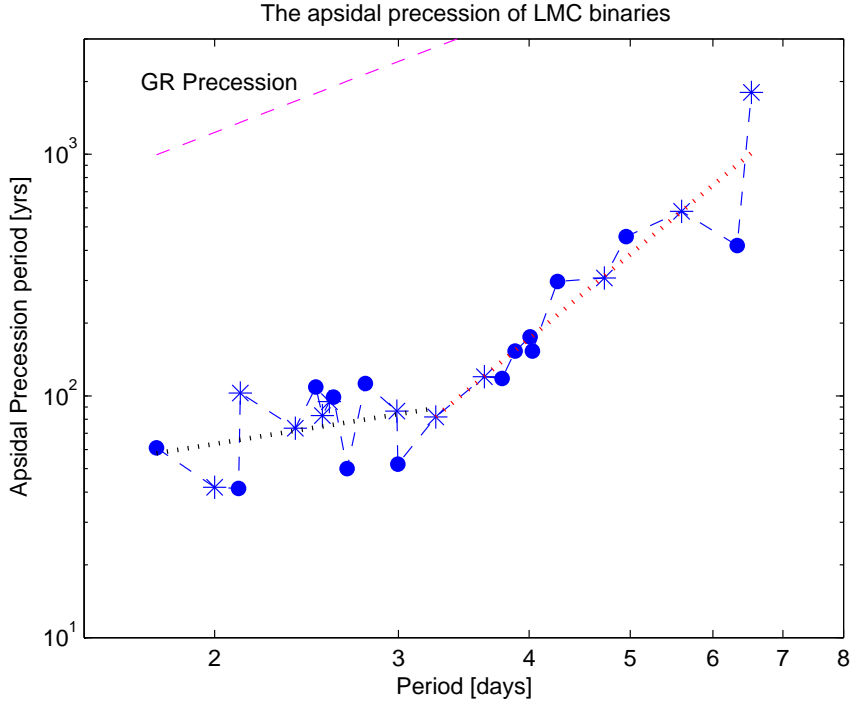} 
\caption{The apsidal precession period as a function of the binary
  period for the LMC binaries. Binaries discussed by Michalska \&
  Pigulski (2005) are denoted by filled circles, and those discussed by 
  Michalska (2007) by $\ast$. A dashed line connects the binaries sorted
  by period. Two straight lines were fitted, one for
  the first 14 short-period binaries, up to a period of 3.2 days, and
  the other for the binaries with longer periods. The expected GR
  precession is also plotted for binaries with total masses of 10
  $M_{\odot}$ and for almost circular orbits.}
\label{apsidal_motion_LMC}
\end{figure}

Unfortunately, the radii and masses of the components of the binaries
of Michalska \& Pigulski and Michalska are not known, and therefore no
comparison between the theory of apsidal motion induced by the tidal
interaction and the observations is possible. However, we can try to
take advantage of the fact that for the OGLE LMC data, most stars are
in a limited magnitude range, and therefore their radii and masses are
not very different from one another. Therefore, the apsidal precession
period should depend mostly on the binary period. To check this
hypothesis we plotted in Figure~\ref{apsidal_motion_LMC} the derived
apsidal period as a function of the binary period. The period range is
between about $2$ and $7$ days, and the apsidal period range is
between $50$ and $2000$ years. For the binaries with periods below $3$
days, the apsidal period shows a linear dependence on the binary
period, as if the ratio $R_1/a$ in Eq. (3.1) is constant, where $R_1$
is the radius of the primary and $a$ is the semi-major axis. For the
binaries with periods longer than $3$ days the apsidal period is
proportional to $ P^{11/3}$, where $P$ is the binary period. Within the
errors this slope is consistent with the exponent of Equation (3.4).

We note that one system, with a period of 6.33 days, has a substantially
shorter precession period than the straight-line fit. This is OGLE LMC
05065201-6825466, the brightest star in the sample of 98 binaries
analysed by Michalska \& Pigulski. This star, with a brightness of $V=
14.04$, probably has a substantially larger radius than the other
stars in Michalska \& Pigulski and Michalska samples, and therefore
might have a shorter apsidal period.

The  Michalska \& Pigulski and Michalska studies comprise the first
systematic attempt to consider the apsidal motion in binaries detected
by large-scaled systematic photometric surveys. We anticipate many more
similar studies in the future.

\subsection{Circularization of eclipsing binaries of the 
SMC and LMC}
\label{subsection_circ_LMC}

A seminal work that used the new sets of binaries to study tidal
interaction was the study of North \& Zahn (\nocite{north2003}2003),
which analysed the detached eclipsing binaries cataloged by Udalski
\etal\ (1998) in the SMC and by Alcock \etal\ (1997) in the LMC.
Because of the magnitude range of the two surveys, all the binaries in
these two samples have early-type primaries, mostly B-type stars
(North \& Zahn 2003). Following Giuricin \etal\ (1984; see a review of
their work in Section~\ref{subsection_early_type}), North \& Zahn
plotted the orbital eccentricity as a function of the fractional radii
of the eclipsing binaries in the SMC and LMC. Like Giuricin \etal\
they tried to find out the cutoff fractional radius, above which all
binaries are circularized. They were also trying to find out whether
there is any difference between the circularization processes in
early-type stars in our Galaxy and those in action in binaries in the
SMC and LMC.

North \& Zahn pointed out that with the precision of the two surveys
it was difficult to derive the orbital eccentricity from the lightcurve
of an eclipsing binary. The phase of the secondary eclipse, which could
be accurately determined, yielded only $e\cos\omega$, where $e$ is the
eccentricity and $\omega$ is the longitude of the periastron. North \&
Zahn overcame this problem by plotting the absolute value of
$e\cos\omega$ instead of the eccentricity itself. As they were
interested only in the statistical features of the sample, and in the
cutoff fractional radius in particular, the fact that only
$e\cos\omega$ was available to the analysis did not hamper their study.

North \& Zahn faced another problem. The given precision of the
lightcurves of the two surveys did not really allow them to derive the
radii of the two components of each binary. One could only get the sum
of the two radii (in terms of the orbital separation), while the ratio
of the two radii was not well determined. North \& Zahn had therefore
to {\it assume} that the two stars have equal radii. Only with the aid of this
assumption could they estimate the primary radius, which is, in fact,
the average of the two radii. With these two modifications, the
analysis of North \& Zahn (2003) yielded results similar to Giuricin
\etal\ (1984). The cutoff fractional radius was about 0.25, with no
statistical difference between the early-type binaries of the Galaxy
and those of SMC and LMC.

In a follow-up paper, North \& Zahn (\nocite{north2004}2004) analysed
many additional eclipsing-binary lightcurves, so that their samples
included 148 binaries in the SMC and 353 binaries in the LMC. They
then found, again, that the critical fractional radius was
in the range $0.24-0.26$, irrespective of mass, surface gravity and
metallicity. This value of the critical radius was consistent with the 
previous studies, and is compatible with the theory of Zahn (1975).

North \& Zahn (2003; 2004) did not use all the binaries detected in
the OGLE LMC data. In fact, Wyrzykowski \etal\ (2004) identified 2580
eclipsing binaries in the LMC OGLE II photometric data alone. However,
it is impractical to manually analyse such a large number of
binaries. One would need an automated analysis to handle such a large
set of binaries. Such algorithms were indeed constructed recently
(e.g., Wyithe \& Wilson \nocite{wyithe2001}2001; Devor \& Charbonneau
\nocite{devor2005}2006). Although the automated codes can err in some
cases, for most of the binaries the algorithms yield the right
solutions, rendering them powerful tools in the
study of the statistical features of the close binary population
(e.g., Mazeh \etal\ 2006).

Following this trend, Tamuz \etal\ (\nocite{tamuz2006}2006)
constructed an automated algorithm, EBAS, to analyze eclipsing binary
lightcurves, an algorithm which is based on the EBOP code (Popper \&
Etzel \nocite{popper1981}1981; Etzel \nocite{etzel1981}1981). The
algorithm was then used by Mazeh \etal\ (\nocite{mazeh2006}2006) to
analyze most of the eclipsing binaries detected by Wyrzykowski \etal\
(2004) in the OGLE LMC data.  Mazeh \etal\ then plotted the
eccentricity as a function of the fractional radius, with two minor
changes relative to the plot of North \& Zahn. First, instead of the
primary fractional radius, they plotted the total fractional radii
$R_{total}=R_1+R_2$ --- where $R_1$ and $R_2$ are the fractional radii
of the primary and secondary, respectively. Second, they specifically
plotted $e\cos\omega$ instead of its absolute value. They found that
the critical total fractional radius is 0.5, consistent with the North \&
Zahn result.

\begin{figure}
\includegraphics[width=13cm]{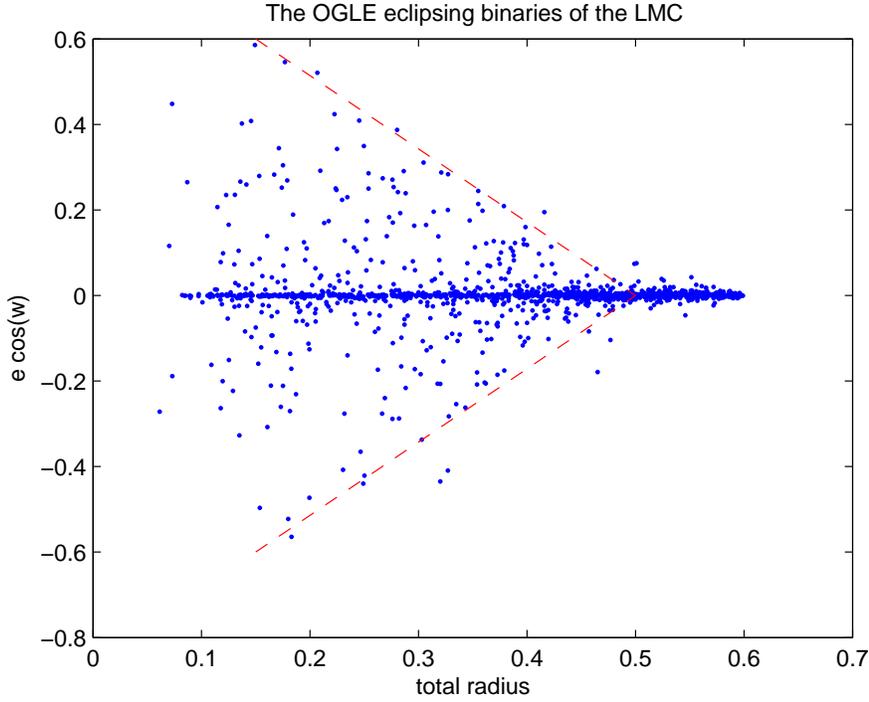}
\caption{The eccentricity, multiplied by $\cos\omega$, of the OGLE
  eclipsing binaries of the LMC as a function of total fractional
  radius. Included are 1335 binaries with $17\leq I \leq 19$. The
  dashed line represents the upper envelope (see text).}
\label{LMC}
\end{figure}

Here we follow Mazeh \etal\ (2006) and plot in Figure~\ref{LMC} $e
\cos \omega$ as a function of the total fractional radius for all 1335
binaries with $19\leq I \leq 17$ analysed by Tamuz \etal\ (2005). We
also plotted a straight line upper envelope, that goes like

\begin{equation}
\pm e\cos\omega=0.857-1.714R_{total} \ ,
\end{equation}
which implies, if we ignore the $\cos\omega$ factor, that the periastron
distance, $1-e$, in terms of the semi-major axis, is 

\begin{equation}
2(R_1+R_2) \stackrel{<}{\sim} 1-e 
\end{equation}
(Mazeh \etal\ 2006). This could mean that if the two stars got too
close at periastron, to the extent that the distance between them was
less than twice their total radius, then circularization processes went into
action and the eccentricity was reduced until the two stars became further
apart. This is not the only interpretation. One can imagine that
binaries were not formed too close together, lest they violate this
relation.  Obviously, the present distribution could be the
caused by the two scenarios --- formation and tidal evolution.

Note that similarly to Giuricin \etal\ (see
Section~\ref{subsection_early_type}), we also find a few binaries with a
primary radius larger than 0.3 which display small but significant
eccentricities. We also tend to attribute these eccentricities to
unseen third distant companions, an effect suggested by Mazeh \&
Shaham (1979; see detailed discussion in Section~\ref{subsection_MS}).

The North \& Zahn (2003; 2004) and Mazeh \etal\ (2006) studies were the
first to use large-scaled photometric surveys to touch upon
circularization interaction.  We anticipate many more, similar studies
in the future.
 
%
\section{Discussion}
%

This paper has attempted to review a large corpus of observational
evidence for tidal interaction in close binary systems, including
short-period extrasolar planets. We first discussed two tidal effects
that can be observed directly --- the ellipsoidal variation and the
apsidal motion. Unfortunately, these two effects have been detected in
the past in only a small number of stars, because the detection
required special photometric observations which involved a
considerable effort. The small number of known binaries with detected
ellipsoidal modulation or apsidal motion allowed for only a limited
confrontation between the theory and the observations. To perform a
comprehensive confrontation we need many more binaries with accurate
measurements of these two effects.

Ellipsoidal modulation was mostly used in the past to study known
spectroscopic binaries (e.g., HD 149420 --- Fekel \& Henry
2006)\nocite{fekel2006}, or binaries with compact objects.  This
situation is being changed dramatically, as ground-based large-scaled
photometric surveys have been operated in the last decade. We already
discussed two such 1-m class exhaustive projects, MACHO (Alcock \etal\
1993\nocite{alcock1993}) and OGLE (e.g., Udalski \etal\
1997\nocite{udalski1997}).  Furthermore, a new wave of observational
effort to search for transiting planets by small, 10-cm class,
telescopes is now at its full thrust (e.g., HATnet --- Bakos \etal\
\nocite{bakos2004}2004; XO --- McCullough \etal\
\nocite{mccullough2005}2005; WASP --- Pollacco \etal\
\nocite{pollacco2006}2006; TrES --- Alonso \etal\
\nocite{alonso2007}2007). These projects are collecting an
unprecedented number of photometric datasets that harbor thousands of
ellipsoidal binaries.

The amplitude of the ellipsoidal modulation can be of the order of a
few percent, and therefore should be easily detected in these large
photometric datasets.  Detecting ellipsoidal variables can become a
tool for discovering new binaries. In such cases, one must distinguish
between the ellipsoidal modulation and other possible variations,
particularly stellar spots and pulsation. For that purpose, one must
rely on the theoretical shape of the modulation and on observations in
different passbands. A first step towards this goal has been performed
already by Derekas \etal\ (2006\nocite{derekas2006}), who identified
ellipsoidal variables in the sample of variable pulsating red giants
within the MACHO data.

The huge datasets are having another effect on the study of close
binaries: they are already revealing a large number of new eclipsing
binaries (e.g., Devor 2005). Therefore the number of known eclipsing
binaries will increase in the next few years by one or two orders of
magnitude. The tens of thousands of new eclipsing binaries will
revolutionize our understanding of close binaries in general and their
statistical features in particular. These datasets will reveal many
other effects of eclipsing binaries, such as apsidal motion and third
distant companions through the time-travel effect. We will then be
well equipped to confront the theory of apsidal motion with observations,
and learn more about the frequency and distribution of triple systems.
We will be able to verify conjectures regarding the correlation
between short-period binaries and distant companions.

The influx of photometric data is about to increase, as more automated
photometric surveys with 1-m class telescopes will set into
action. This includes SkyMaper (Bayliss \& Sackett
2007\nocite{bayliss2007}), Las Cumbres Observatory Global Telescope
(Brown \etal\ \nocite{brown2007}2007), Pan-STARRS (e.g., Holman \etal\
\nocite{holman2007}2007), and finally LSST (e.g., Strauss \etal\
2006\nocite{strauss2006}). One such binary was discovered already by
Blake \etal\ (\nocite{blake2007}2007), using the Sloan Digital Sky
Survey II.  This influx of photometric data will be complemented by
extremely accurate datasets from space missions, which will have an
immediate impact on binary studies. This trend has already been
started by HIPPARCOS (e.g., Jorissen \etal\
\nocite{jorrisen2004}2004), and will be joined by CoRoT (Maceroni and
Ribas \nocite{maceroni2007}2007), Kepler (Koch \etal\
\nocite{koch2007}2007) and GAIA (e.g., Niarchos
2006\nocite{niarchos2006}) missions. For the future of binary studies
in the 'brave new era,' see a review by Guinan \& Engle
(\nocite{guinan2006}2006).

We have discussed the long-term tidal processes of circularization,
synchronization and spin alignment. As was reviewed, binary
circularization has attracted many studies, particularly of the
transition period between circular and eccentric
binaries. Nevertheless, no consensus has been reached with regard to
the theoretical timescale of the tidal processes, and even the
dominant dissipation process behind the circularization is still being
debated.  Synchronization and alignment were studied much less. For
example, the number of eclipsing binaries with observed RM effect by
which we study the spin-orbit alignment is only six, an extremely
small number. Modern spectrographs with efficient digital detectors
can observe many more eclipsing binaries, and the development of tools
to analyze the observations can enable us to accurately derive the
spin-orbit inclination. The results may well be crucial for our
understanding of binary formation, as this will help us estimate the
primordial alignment of binary population. In case binaries are formed
with some spread of spin-orbit inclinations, we should also expect
coeval samples of binaries to exhibit a transition period between
aligned and non-aligned binaries.  Such an observed transition can
serve as another confirmation of the theory of tidal interaction.  It
would be of extreme importance to compare the aligned/non-aligned
transition period with the pseudo-synchronized/non-pseudo-synchronized
and circularized/eccentric transition periods.

An additional feature of the new large accurate photometric datasets
is their long time span, of the order of a few years, that will enable
us to detect minute variations which disclose long-term variations.
The search of Michalska \& Pigulski (2005) and Michalska (2007) for
evidence of apsidal motion discussed above is one example.  The works
of Kubiak \etal\ (\nocite{kubiak2006}2006) and Pilecki \etal\
(\nocite{pilecki2007}2007) are two other examples of studies that
utilized the long time span of the new data sets. Some of the
yet-to-come studies will be able to find eventually direct evidence
for long-term tidal processes in close binaries.

Finally, we have discussed in brief some ideas about tidal evolution
of short-period extrasolar planets. Many of the short-period planets
are being discovered by photometry which detects how these planets
eclipse their parent stars. The transiting planets hold a higher
potential for the study of tidal effects, as we know so much more
about them, including their masses and radii. They can be used as
clean and unique laboratories to study tidal effects, as the
complexity of their systems is reduced because of the small masses and
radii of the planets. As of October 2007, 27 transiting planets are
known, which is about half of the short-period extrasolar planets, the
others have been discovered by radial-velocity measurements. We
anticipate that the balance between the radial-velocity and transiting
planets will change in the near future in favour of the transiting
planets.  The rapidly increasing number of transiting planets will
enable us to study the possible correlation between presumed tidal
effects and the stopping mechanism, the inflated radii of some
planets, synchronization of the stars and planets and their spin-orbit
alignment. A comprehension of these features will help us assess the
role of tidal interaction in the orbital evolution of short-period
extrasolar planets.

%
\section*{Acknowledgements}
%

Special thanks to Jean-Paul Zahn for inviting me to the Oleron summer
school to give a series of lectures on the subject of this
review. This paper emerged from those lectures. I am deeply thankful
for his support and patience during the long period of writing the
manuscript and for his enlightening comments. I am indebted to Soeren
Meibom and Itzhak Goldman for many fruitful discussions and
enlightening comments. Special thanks to Peter Eggleton, Dan Fabrycky,
Scott Gaudi, Mike Lecar, Doug Lin, Andrei Tokovinin and Guillermo
Torres who read an early version of the manuscript and send me
thoughtful comments and suggestions. I wish to heartily thank my
students Avi Shporer, Aviv Ofir and Yevgeny Tsodikovich for technical
help with handling the data and with the plotting.

This work was supported by the Israeli Science Foundation through
grant no. 03/233. I thank the Swiss National Science
Foundation and the Smithsonian Institution for supporting my stay
at the Geneva Observatoire and at the CfA, during which I worked on
this paper.

\bibliographystyle{astron}
\bibliography{57}

\end{document}